\documentclass[onefignum, onetabnum,onealgnum,onethmnum, oneeqnum]{siamonline190516}
\usepackage{comment}
\usepackage{lipsum}
\usepackage[margin=1in, footskip=36pt]{geometry}

\usepackage{endnotes}

\usepackage{titlesec}
\usepackage{titletoc}
\usepackage{pgfplotstable}
\titleclass{\task}{straight}[\section]
\newcounter{task}
\renewcommand{\thetask}{\arabic{task}}
\titleformat{\task}[hang]
    {\normalfont\LARGE\bfseries}{Task \thetask:}{1em}{}

\titleformat*{\task}{\color{header1}\bfseries}

\titlecontents{task}
              [3.8em] 
              {}
              {\contentslabel{2.3em}}
              {\hspace*{-2.3em}}
              {\titlerule*[1pc]{.}\contentspage}

\titlespacing*{\section}{0ex}{1ex}{1ex}
\titlespacing*{\subsection}{0ex}{1ex}{1ex}
\titlespacing*{\subsubsection}{0ex}{1ex}{1ex}
\titlespacing*{\paragraph}{0ex}{1ex}{1ex}
\titlespacing*{\subparagraph}{0pt}{1ex}{1ex}
\titlespacing*{\task}{0em}{1ex}{1ex}

\usepackage[sort&compress,comma,square,numbers]{natbib}

\usepackage{amsfonts,amsmath,amssymb}
\usepackage{bbm}

\usepackage{multicol}
\usepackage{enumitem}
\setlist[enumerate]{wide, labelindent=1cm,  noitemsep}
\setlist[itemize]{noitemsep}
\setlist[description]{noitemsep}

\usepackage{hhline}

\usepackage{todonotes}
\RequirePackage{ulem}

\definecolor{tableheadcolor}{gray}{0.92}
\newcommand{\topline}{ %
        \arrayrulecolor{blue1}\specialrule{0.1em}{\abovetopsep}{0pt}%
        \arrayrulecolor{tableheadcolor}\specialrule{\belowrulesep}{0pt}{0pt}%
        \arrayrulecolor{blue1}}
\newcommand{\midtopline}{ %
        \arrayrulecolor{tableheadcolor}\specialrule{\aboverulesep}{0pt}{0pt}%
        \arrayrulecolor{blue1}\specialrule{\lightrulewidth}{0pt}{0pt}%
        \arrayrulecolor{white}\specialrule{\belowrulesep}{0pt}{0pt}%
        \arrayrulecolor{blue1}}
\newcommand{\bottomline}{ %
        \arrayrulecolor{white}\specialrule{\aboverulesep}{0pt}{0pt}%
        \arrayrulecolor{blue1} %
        \specialrule{\heavyrulewidth}{0pt}{\belowbottomsep}}%

\pgfplotstableset{normal/.style ={%
        header=true,
        string type,
        font=\small,
    	column type={p{.092\textwidth}},
        every head row/.style={
            before row={\topline\rowcolor{tableheadcolor}},
            after row={\midtopline}
        },
        every odd row/.style={
            before row={\rowcolor{blue1}}
        },
        every last row/.style={
            after row=\bottomline
        },
        col sep=&,
        row sep=\midrule
    }
}

\usepackage{multirow}

\usepackage{booktabs}
\usepackage{fancyhdr}
\usepackage{fullpage}

\RequirePackage{titlesec}
\RequirePackage[font={footnotesize},{singlelinecheck=false}]{caption}
\usepackage[T1]{fontenc}
\usepackage[scaled]{helvet}
\usepackage[scaled=1.10, med]{zlmtt}

\usepackage{algorithm}
\usepackage{algpseudocode}

\providecommand{\sct}[1]{{\sc \texttt{#1}}}

\newcommand{\Mgc}{\sct{Mgc}}

\newcommand{\Dcorr}{\sct{Dcorr}}

\providecommand{\ve}[1]{\boldsymbol{#1}}
\providecommand{\ma}[1]{\boldsymbol{#1}}

\providecommand{\norm}[1]{\ensuremath{\left \lVert#1 \right  \rVert}}

\newcommand{\trace}[1]{{\ensuremath{\operatorname{tr}\!\left(#1\right)}}}           
\providecommand{\mc}[1]{\mathcal{#1}}

\providecommand{\mb}[1]{\boldsymbol{#1}}

\providecommand{\mt}[1]{\widetilde{#1}}

\providecommand{\mv}[1]{\vec{#1}}
\providecommand{\mh}[1]{\hat{#1}}

\providecommand{\mtb}[1]{\widetilde{\boldsymbol{#1}}}

\newcommand{\conv}{\rightarrow}

\newcommand{\mcE}{\mathcal{E}}

\newcommand{\PP}{\mathbb{P}}

\DeclareMathOperator{\R}{R} 
\DeclareMathOperator{\Y}{\mathbf{Y}}

\DeclareMathOperator{\X}{\mathbf{X}}

\newcommand{\bth}{\ve{\theta}}

\newcommand{\thetn}{\ve{\theta}}
\newcommand{\thet}{\thetn}

\floatname{algorithm}{Algorithm}
  
\usepackage[symbol]{footmisc}

\usepackage{ericmath}
\usepackage[createShortEnv]{proofs}

\definecolor{fig1green}{rgb}{0.01, .7, 0.1}
\definecolor{fig1orange}{rgb}{0.7, 0.34, 0.005}

\newcommand{\anova}{\sct{anova}}
\newcommand{\vm}{\sct{VM}}
\newcommand{\hsic}{\sct{Hsic}}
\newcommand{\cdcorr}{\sct{cDcorr}}
\newcommand{\pdcorr}{\sct{pDcorr}}

\newcommand{\manova}{\sct{manova}}
\newcommand{\cmanova}{\sct{cManova}}

\newcommand{\kcd}{\sct{kernelCDTest}}
\newcommand{\causal}{\sct{Causal}}
\newcommand{\gcm}{\sct{GCM}}
\newcommand{\rcit}{\sct{RCIT}}
\newcommand{\rcot}{\sct{RCoT}}
\newcommand{\ccdcorr}{\causal~\cdcorr}

\title{
Learning sources of variability from high-dimensional observational studies
}

\author{
    Eric W.~Bridgeford$^{1,\dagger}$, 
    Jaewon Chung$^1$,
    Brian Gilbert$^1$,
    Sambit Panda$^1$,
    Adam Li$^3$,
    Cencheng Shen$^2$,
    Alexandra Badea$^4$,
    Brian Caffo$^1$,
    Joshua T.~Vogelstein$^1$.
    \thanks{
     $^1$ Johns Hopkins University, $^2$ University of Delaware, $^3$ Columbia University, $^4$ Duke University
    $^\dagger$ Corresponding author:
      Eric W.~Bridgeford (\email{ebridge2@jhu.edu}).}
}

\makeatletter
\def\thanks#1{\protected@xdef\@thanks{\@thanks
        \protect\footnotetext{#1}}}
\makeatother
\makeatletter
\renewcommand\@biblabel[1]{#1.}
\makeatother
\begin{document}

\maketitle

\begin{abstract}
Causal inference studies whether the presence of a variable influences an observed outcome. As measured by quantities such as the ``average treatment effect,'' this paradigm is employed across numerous biological fields, from vaccine and drug development to policy interventions. Unfortunately, the majority of these methods are often limited to univariate outcomes. Our work generalizes causal estimands to outcomes with any number of dimensions or any measurable space, and formulates traditional causal estimands for nominal variables as causal discrepancy tests. We propose a simple framework under which universally consistent conditional independence tests are universally consistent causal discrepancy tests. Numerical experiments illustrate that our method, \ccdcorr, leads to improvements in both finite sample validity and power when compared to existing strategies when the assumptions of this framework are violated. 
Our methods are all open source and available at \href{github.com/ebridge2/cdcorr}{github.com/ebridge2/cdcorr}.
\end{abstract}

%
%
\section{Introduction}

Many of the earliest developments in statistics have focused on differentiating sources of variability in scientific data. In one of the most impactful works of statistics, \citet{Fisher1935} pioneered the concept of the \anova, a test for determining whether two groups of outcomes fundamentally differ from one another (the \textit{two-sample test}). This early work was made possible for univariate outcomes through assumptions about how the outcomes could differ (that is, each of the two samples of outcomes were $iid$ within-group) and strong distributional assumptions. To overcome the former limitation, early efforts focused on the development of regression techniques to account for factors which influenced the distribution of each group of outcomes (so called `covariates'), loosening $iid$ assumptions to independence given the covariate \cite{Agresti2015Feb,McCullagh1989Aug}. Similarly, these regression techniques allowed further generalizations of \anova~ to the case where the outcomes have more than two groups, facilitating a so-called conditional \anova \cite{McCullagh1989Aug,Agresti2015Feb}.

More recently, techniques have been developed to address other limitations of \anova. \manova, or multivariate \anova, extends the idea of determining differences in distribution from univariate to multivariate data via leveraging parametric Gaussian assumptions \cite{Pillai1976,Pillai1977,Rao1951}. The assumption of multivariate Gaussianity have been further loosened via independence tests \cite{Szekely2007Dec}, which can be combined with $iid$ assumptions to produce asymptotically consistent tests on arbitrary metric spaces. It has been illustrated that these tests can be augmented \cite{Panda2019Oct} to facilitate non-parametric $K$-sample testing. Correspondingly, numerous conditional independence tests have been proposed \cite{Strobl2019Mar,Wang2015,Shah2018Apr} which therefore enable non-parametric conditional $K$-sample testing. Unfortunately, these more complicated strategies are not without limitations. \citet{Shah2018Apr} demonstrate that there is no conditional independence test which can achieve both sensitivity and specificity without additional assumptions placed on the joint distribution of the outcomes (for each group) and the covariates. 

These developments in the statistical hypothesis testing literature have been juxtaposed by efforts to determine whether a predictor actually `causes' an effect on an outcome in the presence of other covariates \cite{Rubin1974,Athey2017May}. In the simplest possible case where the potential predictor is binary (it is either present, or it is not present), the predictor is often known as a `treatment,' and the causal estimand is known as an `average treatment effect' \cite{Rosenbaum1983Apr,Rosenbaum1984Mar} or a `conditional average treatment effect'~\cite{Robins1986Jan} (the effect of the treatment on the outcome, depending on the value taken by the covariates). These methods can be understood to address the limitations posed by \citet{Shah2018Apr} by placing additional assumptions on the joint distribution via a `potential outcome' or `counterfactual' framework, where the causal effect is measured in terms of differences (usually in the expectation) of the potential outcomes \cite{Hernán2006,Pearl2009Jan,Pearl2010Jul}. 

An effort to harmonize the conditional independence testing and causal literature has been addressed by \textit{multivariate causal discrepancy testing}, which focuses on the question of determining whether two potential outcome distributions differ \cite{Park2021Jul}. It is unclear the extent to which these techniques generalize when the samples do not overlap in terms of upstream covariates (more specifically, the \textit{confounders}; so called ``covariate imbalance''), which is an  common data presentation in observational settings. Informally, it is unclear how robust these techniques are to the situation when the underlying causal assumptions fail, or are poorly reflected in the data sample. Moreover, there is little insight into formally characterizing causal discrepancy testing for more than two groups, which leaves a logical disconnect for many settings. Finally, it is unclear how sensitive and specific these tests are under non-linearities and non-monotonicities, which presents a substantial hurdle to their utility in high-dimensional settings.

Inspired by \citet{Park2021Jul} and \citet{Bridgeford2022Oct}, we introduce practical estimands which generalize the concepts of causal effects to data of arbitrary distribution (so long as realizations are measurable) and any number of treatment groups. 
First, we prove that, under general assumptions, any conditional independence test can be used to produce a $K$-sample causal conditional discrepancy test in both randomized clinical studies and observational studies. This result directly unifies the rich field of independence testing with causal discrepancy testing, and provides explicit theoretical motivation for the use of independence tests in \textit{causal structure learning} \cite{Robins2000Sep} with nominal treatment variables. Second, we illustrate additional assumptions needed to generalize these results to unconditional potential outcomes, drawing a parallel between causal conditional discrepancy testing with causal unconditional discrepancy testing in the same manner as the parallel between the average treatment effect and the conditional average treatment effect. Third, we propose simple and practical procedures for augmenting existing approaches for conditional independence testing to better suit the conditional discrepancy testing regime. 

Our proposed strategy, \ccdcorr, achieves substantial improvements in finite-sample validity and finite-sample power over existing approaches that are typically used for conditional discrepancy testing and conditional independence testing in the case of nominally-grouped data across regimes with both low and high degrees of covariate imbalance. We explore the values of this perspective across a range of simulations in both low and high-dimensional regimes, and show that our augmentations to existing conditional independence tests facilitate principled causal inference across a variety of multivariate contexts in which other techniques do not generalize, including non-linearities and non-monotonicities, higher moment differences across groups, and multi-group settings. Together, we believe that these results suggest the value of harmonizing causal perspectives with cutting-edge developments in non-parametric statistics. 

\section{Preliminaries}

Throughout this work, we use the notation defined in Table \ref{tab:causal:prelim}.
\begin{table}[]
    \centering
    \begin{tabular}{|c|c|}
        \hline
         Symbol & Meaning \\
         \hline
         $\mathbf u_i$ & a random variable \\
         $u_i$ & a realization of a random variable $\mathbf u_i$ \\
         $\mathcal U$ & a space defining the values taken by realizations; e.g., $u_i \in \mathcal U$ \\
         $[N]$ & shorthand for the set $\{1, \hdots, N\}$ \\
         $\mathbf y_i$ & a random variable representing an outcome \\
         $\mathbf x_i$ & a random variable representing a covariate \\
         $\mathbf t_i$ & a random variable denoting group/treatment \\
         $F_{\mathbf u_i, \mathbf w_i}$ & joint distribution of $(\mathbf u_i, \mathbf w_i)$ \\
         $f_{\mathbf u_i, \mathbf w_i}(u, w)$ & joint density evaluated at $(u, v)$ \\
         $F_{\mathbf u_i | \mathbf w_i}$ & distribution of $\mathbf u_i$ conditional on $\mathbf w_i$ \\
         $F_{\mathbf u_i | w}$ & distribution of $\mathbf u_i$ conditional on the event $\mathbf w_i = w$ \\
         $\prob{\mathbf u_i = u}$ & probability mass for the event where $\mathbf u_i = u$ \\
         $\mathbf y_i(k)$ & potential outcome for an individual $i$ in group $k$ \\
         \hline
    \end{tabular}
    \caption{The notation table used within this work.}
    \label{tab:causal:prelim}
\end{table}

\subsection*{From \anova~to conditional $K$-sample testing}

Some of the earliest works that yielded the development of modern statistical inference addressed the $2$-sample testing problem. In its simplest form, the outcome $\mathbf y_i$ takes values $y \in \mathbb R$ and the group $\mathbf t_i$ takes values $t \in [2]$. The $2$-sample testing problem is defined as a test of:
\begin{align*}
    H_0 : F_{\mathbf y_i | 1} = F_{\mathbf y_i | 2}\text{ against }H_A : F_{\mathbf y_i | 1} \neq F_{\mathbf y_i | 2}.
\end{align*}
In the case where $F_{\mathbf y_i | t}$ are $\Norm{\mu_t, \sigma^2}$ and each $\mathbf y_i$ are independent samples, this problem is known as the one-way \anova \cite{Fisher1935}, and can be addressed via the $F$ test \cite{Agresti2015Feb,McCullagh1989Aug}. 

This was later relaxed via the development of the development of the conditional \anova, wherein the test can be sufficiently generalized to $K$-samples, and can incorporate other covariates via a likelihood ratio test \cite{Agresti2015Feb,McCullagh1989Aug}. In this case, we are interested in the $K$-sample conditional discrepancy, defined in Definition \ref{def:cond_kst}.
\begin{definition}[$K$-sample conditional discrepancy]
For outcomes $\mathbf y_i$ taking values $y \in \mathcal Y$, a grouping variable $\mathbf t_i$ taking values $t \in [K]$, and a set of covariates $\mathbf x_i$ taking values $x \in \mathcal X$, a $K$-sample conditional discrepancy exists if for some $k, l \in [K]$ and $x \in \mathcal X$, then:
\begin{align*}
    F_{\mathbf y_i | k, x} \neq F_{\mathbf y_i | l, x}.
\end{align*}
\label{def:cond_kst}
\end{definition}
Intuitively, a $K$-sample conditional discrepancy exists if, conditional on the covariates $x$, there is a difference in the outcome distributions for some pair of groups $k$ and $l$ conditional on the covariates $x$. A natural test can be developed with:
\begin{align*}
    H_0 &: \text{a conditional discrepancy does not exist for any $k, l, x$} \\
    H_A &: \text{a conditional discrepancy exists}.
    \numberthis\label{eqn:cond_kst_discrep}
\end{align*}
The $K$-sample conditional discrepancy can be intuited via linear regression, where assuming that $\pmb \epsilon_i  \distas{iid} \Norm{0, \sigma^2}$:
\begin{align*}
    \mathbf y_i = \mu_{\mathbf t_i}(\mathbf x_i) + \pmb \epsilon_{i},
\end{align*}
In the case where $\mu_{\mathbf t_i}(\mathbf x_i) = \mu_{\mathbf t_i} + f(\mathbf x_i) + g(\mathbf x_i, \mathbf t_i)$ (a model with a group-specific offset, a covariate-specific term delineated by $f$, and an interaction term delineated by $g$) and the functions $f$ and $g$ are assumed to be known, this can be tested directly via the likelihood ratio test \cite{Agresti2015Feb}. With this linear regression intuition in mind, the interpretation of such a test is very similar to that of the one-way \anova, and the hypothesis simplifies to a test of whether the group means $\mu_{t}$ and the interactions are equal for all groups against the alternative that for some pair of groups, they are unequal.

This approach was further relaxed by the conditional \manova, wherein we instead suppose that $\vec{\pmb \epsilon}_i \distas{iid} \Norm{\vec 0, \Sigma}$, and we allow $\mathbf y_i$ to take values $\vec y \in \mathbb R^{P}$, where for each dimension $p$, we model:
\begin{align*}
    \mathbf y_{ip} = \mu_{\mathbf t_i p} + f_p(\mathbf x_i) + g_p(\mathbf x_i, \mathbf t_i) + \pmb \epsilon_{ip}.
\end{align*}
A suitable test can be developed using this strategy via \cmanova \cite{Pillai1976,Pillai1977,Rao1951,Jobson1992}, detailed in Appendix \ref{app:methods:cmanova}, when the functions $f$ and $g$ are known (optionally, additional parametric assumptions may be placed on $\Sigma$ to yield an identifiable solution to the linear model \cite{McCullagh1989Aug,Jobson1992}). Conclusions do not generally apply without Gaussian assumptions, and the technique cannot be applied to high-dimensional datasets without additional parametric assumptions or regularization \cite{Cai2014Oct}.

A closely related problem to the $K$-sample conditional discrepancy problem is the conditional independence testing problem. It is framed as follows: we observe samples $(\vec y_i, \vec v_i, \vec x_i) \in \mathbb R^p \times \mathbb R^q \times \mathbb R^r$ for $i \in [n]$. We suppose the existence of three random variables $\vec{\mathbf y}_i$, $\vec{\mathbf w}_i$, and $\vec{\mathbf y}_i$, where $(\vec{\mathbf y}_i, \vec{\mathbf v}_i, \vec{\mathbf x}_i)$ are sampled independently and identically from $F_{\mathbf y_i,\mathbf v_i,\mathbf x_i}$. The two random variables $\vec{\mathbf y}_i$ and $\vec{\mathbf v}_i$ are independent conditionally on $\vec{\mathbf x}_i$ if and only if $F_{\mathbf y_i, \mathbf v_i | \mathbf x_i} = F_{\mathbf y_i | \mathbf x_i}F_{\mathbf v_i|\mathbf x_i}$. So, the conditional independence testing problem can be stated as:
\begin{align}
    H_0 : F_{\mathbf y_i, \mathbf v_i|\mathbf x_i} = F_{\mathbf y_i|\mathbf x_i}F_{\mathbf v_i|\mathbf x_i} \text{ against }H_A : F_{\mathbf y_i, \mathbf v_i|\mathbf x_i} \neq F_{\mathbf y_i|\mathbf x_i}F_{\mathbf v_i|\mathbf x_i}.
    \label{eqn:hypo_cond_ind}
\end{align}

Under general assumptions, consistent conditional independence tests of Equation \eqref{eqn:hypo_cond_ind} are consistent $k$-sample conditional discrepancy tests from Equation \eqref{eqn:cond_kst_discrep}, as explained in Remark \ref{rem:ind_test_kst}.

\begin{remark}[Consistent independence testing and consistent $k$-sample conditional discrepancy testing]
Suppose the setup described in \ref{setup}, and let $\mathbf v_i$ be a random $K$-dimensional vector, where for each $t \in [K]$:
\begin{align*}
    \mathbf v_{it} = g(\mathbf t_i) = \begin{cases}
        1 & \mathbf t_i = t \\
        0 & \mathbf t_i \neq t
    \end{cases}.
\end{align*}
Then for any $k, l \in [K]$ and $x \in \mathcal X$, $F_{\mathbf y_i | k, x} \neq F_{\mathbf y_i | l, x}$ if and only if $F_{\mathbf y_i, \mathbf v_i | \mathbf x_i = x} \neq F_{\mathbf y_i | \mathbf x_i = x}F_{\mathbf v_i | \mathbf x_i = x}$.
\label{rem:ind_test_kst}
\end{remark}

This trivial result proven explicitly by direct application of the main result from \citet{Panda2019Oct} allows us to tie together $k$-sample conditional discrepancy testing with conditional independence testing, and can therefore be used to construct a relaxation of the assumptions inherent in the \cmanova~framework (Gaussianity, and choice of the functions $f_p$ and $g_p$ for all $p$). The Generalized Covariance Measure (\gcm) \cite{Shah2018Apr} addresses this problem using a regression of $\mathbf y_i$ onto $\mathbf v_i$ conditional on $\mathbf x_i$. This strategy instead investigates vanishing correlation, a normalized covariance between the scaled residuals and $(\mathbf v_i, \mathbf x_i)$ \cite{Li2019Dec}. Further, \gcm~flexibly extends the intuition of \cmanova~to higher-dimensional settings and achieves consistency outside of gaussian contexts \cite{Shah2018Apr}.

Derivatives of the Hilbert-Schmidt Information Criterion (\hsic) leverage normalized conditional cross-covariance operators (such as \kcd) on reproducing kernel Hilbert spaces (RKHSs) \cite{Park2021Jul}, but are limited to $2$-class settings. Other generalizations leveraging RKHSs have been proposed, such as the Randomized Conditional Independence Test (\rcit) and the Randomized Correlation Test (\rcot) \cite{Strobl2019Mar}, but the finite-sample performance of these techniques in $k$-sample regimes and when the positivity criterion is not ensured \textit{a priori} are unknown. Further, two generalizations of Energy statistics, the conditional distance correlation (\cdcorr) and the partial distance correlation (\pdcorr) have been developed to subvert these limitations under the growing distance correlation framework \cite{Szekely2007Dec}. \pdcorr~provides numerous intuitive and computational advantages similar to \Dcorr \cite{Szekely2007Dec}, but unfortunately is not a dependence measure \cite{Szekely2014Dec}. \cdcorr~provides a test which has shown high testing power under a range of dependence structures, particularly when the relationship between the two random variables given the third is non-monotonic or non-linear \cite{Wang2015}. 

\subsection*{From average treatment effects to causal conditional discrepancies}

Juxtaposed by the developments in the conditional testing literature, the average treatment effect \cite{Rubin1974,Athey2017May} has long formed the backbone of many investigations in causal inference. We obtain the observed data $(y_i, t_i,  x_i)$, where $y_i \in \mathbb R$ is the outcome, $t_i \in \{1, 2\}$ is the (binary) treatment/intervention of interest (either treated or untreated), and $ x_i$ are the vector of baseline covariates (potential confounders), for individuals $i \in [n]$. To investigate this problem, we assume the existence of three random variables, $(\mathbf y_i, \mathbf t_i, {\mathbf x}_i)$, which are sampled independently and identically from some unknown data-generating distribution $F_{\mathbf y_i, \mathbf t_i, \mathbf x_i}$. Additionally, we assume the existence of two counterfactual random variables, $\mathbf y_i(1)$ and $\mathbf y_i(2)$, which represent the \textit{potential} outcomes under the two possible treatments. Conceptually, the theory of causal inference rests on the \textbf{consistency} assumption, which asserts that there is a single \textit{version} of each treatment level; e.g., if $\mathbf t_i = t$, then $\mathbf y_i = \mathbf y_i(t)$ \cite{Cole2009Jan}. Under this framework, the observed outcome is:
\begin{align*}
    \mathbf y_i &= \mathbf y_i(1) \indicator{\mathbf t_i = 1} + \mathbf y_i(2) \indicator{\mathbf t_i = 2},
\end{align*}
where only one of the potential outcomes will actually be realized in the observed data. The average treatment effect ($ATE$) is given by:
\begin{align}
    \gamma \triangleq ATE = \expect{\mathbf y_i(2) - \mathbf y_i(1)} = \expect{\mathbf y_i(2)} - \expect{\mathbf y_i(1)}
\end{align}
To test whether there is an $ATE$ implies the following hypothesis test:
\begin{align}
    H_0 : \gamma = 0 \text{ against }H_A: \gamma \neq 0.
\end{align}

The most obvious issue regarding the ATE is that, in practice, we observe realizations of $\mathbf y_i$ (the observed data), and \textit{not} $\mathbf y_i(t)$ (the counterfactual data); this is known as the ``fundamental problem of causal inference.'' When is $\expect{\mathbf y_i(t)}$ identifiable from the observed data, and how do we identify it?

The identifiability of $\expect{\mathbf y_i(t)}$ is ensured by the \textit{ignorability}, \textit{consistency}, \textit{positivity}, and the \textit{no interference} constraints. \textbf{Ignorability} is said to hold provided that $(\mathbf y_i(2), \mathbf y_i(1)) \indep \mathbf t_i \cond {\mathbf x}_i$; that is, the treatment is independent with respect to the observed baseline covariates. This condition will hold if the study executes it by design (such as in a perfect randomized trial) or all confounders have been recorded with the baseline covariates \cite{Holland1986}. The \textit{consistency} assumption is described above. \textbf{Positivity} holds if, for each possible treatment $t$, $Pr(\mathbf t_i = t \cond  {\mathbf x}_i =  x) > 0$ for any $ x$ in the support of ${\mathbf x}_i$ \cite{Rosenbaum1985,Rosenbaum2010}. Conceptually, any possible individual with a given covariate level \textit{could} have been observed in either the treated or untreated group. \textbf{No interference} asserts that there is no impact between the treatment assignments of \textit{other} participants on the potential outcomes of a given participant; this allows $\mathbf y_i(k)$ to be well-defined without reference to other individuals' treatment assignments \cite{Hernán2006}. When the consistency and no interference assumptions hold, these two assumptions are collectively referred to as the \textbf{Stable-Unit Treatment Value Assumption} (SUTVA) \cite{Hernán2006,Imbens2015,Rubin1980}. 

Under the $G$-computation formula \cite{Robins1986Jan}, if these assumptions hold, then:
\begin{align*}
    \expect{\mathbf y_i(t)} &= \expect{\expect{\mathbf y_i \cond \mathbf t_i = t,  {\mathbf x}_i}},
\end{align*}
and the ATE can be expressed as:
\begin{align}
    \gamma &= \expect{\expect{\mathbf y_i \cond \mathbf t_i = 2,  {\mathbf x}_i}} - \expect{ \expect{\mathbf y_i \cond \mathbf t_i = 1, {\mathbf x}_i}}.
    \label{eqn:g_comp}
\end{align}

As a general measure of treatment effects, the average treatment effect, as defined above, is rather limiting. First, treatment effects aren't necessarily constant across all individuals. One could conceptualize an intervention which has a strongly positive effect on younger people, but has a negative effect on older people. The average treatment effect ends up "averaging away" the heterogeneous effect of treatment on younger and older people, with the average treatment effect ending up being zero. This has been overcome by study of conditional (on baseline covariates, such as age) average treatment effects (the $CATE$) \cite{hahn1998role}: 
\begin{align*}
    \gamma_x \triangleq CATE(x) = \expect{\mathbf y_i(2) - \mathbf y_i(1) \cond \mathbf x_i = x} = \expect{\mathbf y_i(2) \cond \mathbf x_i = x} - \expect{\mathbf y_i(1) \cond \mathbf x_i = x}
\end{align*}
and a relevant test is whether, for each possible $x$:
\begin{align}
    H_0 : \gamma_x = 0 \text{ against }H_A: \gamma_x \neq 0,
    \label{eqn:hypo_cate}
\end{align}
but these too are not without limitations. These limiting definitions of treatment effects are well-defined for multivariate data, in that two multivariate random variables can differ in expectation (or not). However, treatments may impact outcomes beyond simple differences in expectation, such as differences in higher order moments. While one could augment the outcome of interest to be other useful functions such as the square of the outcome, it is unclear in practice how to search over this space of measurable functions to better characterize treatment effects. The setup that we follow for the remainder of this work is noted in Setup \ref{setup}.

\begin{setup}[Causal]
We obtain the observed data $(y_i, t_i, x_i)$, where $y_i \in \mathcal Y$ is the outcome, $t_i \in [T]$ is the nominal treatment, and $x_i \in \mathcal X$ are a collection of baseline covariates (potential confounders), for $i \in [n]$ individuals. We assume that the tuple $(y_i, t_i, x_i)$ is a realization of the random tuple $(\mathbf y_i, \mathbf t_i, \mathbf x_i)$, where:
\begin{enumerate}
    \item $\mathbf y_i$ is the $\mathcal Y$-valued random outcome, where $(\mathcal Y, \delta_y)$ is a metric space, 
    \item $\mathbf t_i$ is the nominal treatment which is a $[T]$-valued random variable, 
    \item $\mathbf x_i$ are the $\mathcal X$-valued random baseline covariates, where for all $x \in \mathcal X$, $f_{\mathbf x_i}(x) > 0$,
    \item For each individual, the treatment assignment mechanism is \textbf{consistent}, where the counterfactual random variables $\mathbf y_i(k)$ represent the potential outcome under treatment $k$, with distribution $F_{\mathbf y_i(k)}$.
    The outcome is:
    \begin{align*}
        \mathbf y_i &= \sum_{k = 1}^K \mathbf y_i(k) \indicator{\mathbf t_i = k}.
    \end{align*}
\end{enumerate}
We assume that the tuples $(\mathbf y_i, \mathbf t_i, \mathbf x_i)$ are sampled independently and identically from some unknown data-generating distribution $F_{\mathbf y_i, \mathbf t_i, \mathbf x_i}$. Note that the potential outcome $\mathbf y_i(k)$ is \textit{not} a function of the treatment assignments for any other individuals $j \neq i$, implying that the \textbf{no interference} criterion is satisfied.
\label{setup}
\end{setup}

A natural generalization of the hypothesis given in Equation \eqref{eqn:hypo_cate} for the case where $K=2$ to data of arbitrary distribution is the causal conditional discrepancy test, first explored explicitly by \citet{Park2021Jul}:
\begin{align}
    H_0 : F_{\mathbf y_i(1)|\mathbf x_i = x} = F_{\mathbf y_i(2)|\mathbf x_i = x} \text{for all $x \in \mathcal X$    against    }H_A :  F_{\mathbf y_i(1)|\mathbf x_i = x} \neq F_{\mathbf y_i(2)|\mathbf x_i = x}\text{ for some $x \in \mathcal X$}.
    \label{eqn:hypo_causal_2s}
\end{align}
That this hypothesis can be relaxed to arbitrary $K$-sample tests, and can be generalized and tested via augmentations of any conditional independence test, serves as the motivation for this work.

\section{Theory}

\subsection{From conditional $K$-sample tests to causal conditional discrepancy tests}

We can further relax this to the case where we have arbitrarily many treatment levels, by a modification of the definition presented by \citet{Park2021Jul}:
\begin{definition}[$K$-sample causal conditional discrepancy]
Suppose the setup described in Setup \ref{setup}. A $K$-sample causal conditional discrepancy exists if for any $k, l \in [K]$ and for any $x \in \mathcal X$:
\begin{align*}
    F_{\mathbf y_i(k)|\mathbf x_i = x} \neq F_{\mathbf y_i(l) | \mathbf x_i = x}.
\end{align*}
\label{def:CoDiCE}
\end{definition}
We explicitly incorporate the language \textit{causal} conditional discrepancy to emphasize explicitly that this effect represents a conditional (on the covariates) discrepancy in the \textit{potential} outcome distributions under nominal treatments. In this case for $K=2$, this definition is equivalent to the definition of a Conditional Distributional Treatment Effect (CoDiTE) given by \citet{Park2021Jul}. A suitable hypothesis test for a $K$-sample causal conditional discrepancy is a $K$-sample causal discrepancy test.
\begin{definition}[$K$-sample causal conditional discrepancy test]
    \label{def:hypo_causal_kst}
Suppose the setup in Setup \ref{setup}, where the nominal treatment variable $\mathbf t_i$ is a $[K]$-valued random variable where $K \geq 2$. A $K$-sample causal discrepancy test is:
\begin{align*}
    H_0 : F_{\mathbf y_i(k)|\mathbf x_i = x} = F_{\mathbf y_i(l)|\mathbf x_i = x}\text{ for all $k, l, x$   against     }H_A : F_{\mathbf y_i(k)|\mathbf x_i = x} \neq F_{\mathbf y_i(l)|\mathbf x_i = x}\text{ for some $k, l, x$.}
\end{align*}
\end{definition}
As it was for the CATE, we do not observe realizations of $\mathbf y_i(k)$ in practice, as these outcomes are only \textit{potential} outcomes under the treatment. This means that the hypothesis test in Definition \ref{def:hypo_causal_kst} is not directly testable in practice without additional assumptions.

Unlike the $K$-sample causal discrepancy test, which tests for discrepancies in the conditional (on only the covariates) distributions of the \textit{potential} outcomes, the $K$-sample conditional discrepancy test implied by Definition \ref{def:cond_kst} in Equation \eqref{eqn:cond_kst_discrep} for discrepancies in the conditional (on \textit{both} the covariates and the nominal treatment assignment $\mathbf t_i$) distributions of the \textit{realized} outcomes $\mathbf y_i$. Under causal assumptions, the $K$-sample conditional discrepancy test is a $K$-sample causal discrepancy test.

\begin{lemmaE}[$K$-sample conditional and causal discrepancy testing equivalence]
Assume the setup described in \ref{setup}. Further, suppose that:
\begin{enumerate}
    \item The treatment assignment is ignorable: $\parens*{\mathbf y_i(1), ..., \mathbf y_i(K)} \indep \mathbf t_i \cond \mathbf x_i$, and
    \item The treatment assignments are positive for all levels of the covariates: $\prob{\mathbf t_i = k \cond \mathbf x_i = x} > 0$ for any $x \in \mathcal X$.
\end{enumerate}
Then a consistent test of Equation \eqref{eqn:cond_kst_discrep} for a $K$-sample conditional discrepancy is equivalent to a $K$-sample causal conditional discrepancy test in Definition \ref{def:hypo_causal_kst}.
\label{thm:CoDiCE}
\end{lemmaE}
\begin{proofE}
Recall that for Equation \eqref{eqn:cond_kst_discrep} and Definition \ref{def:hypo_causal_kst}, that since the density fully determines a distribution, that a difference in distribution exists $\iff$ a difference in the densities exists.

For any $k, l \in [K]$ and $x \in \mathcal X$, note that Equation \ref{def:hypo_causal_kst} can be expressed in terms of the relevant densities from Equation \eqref{eqn:cond_kst_discrep}:
\begin{align*}
    f_{\mathbf y_i(k)|\mathbf x_i = x}(y) - f_{\mathbf y_i(l)|\mathbf x_i = x}(y) &= \sum_{t \in \mathcal [K]}\parens*{f_{\mathbf y_i(k)| t, x}(y) - f_{\mathbf y_i(l)| t, x}(y)}\prob{\mathbf t_i = t | \mathbf x_i = x}.
\end{align*}
By ignorability and consistency, $f_{\mathbf y_i(k) | t, x} = f_{\mathbf y_i | k, x}$ for all $t \in [K]$, so:
\begin{align*}
    f_{\mathbf y_i(k)|\mathbf x_i = x}(y) - f_{\mathbf y_i(l)|\mathbf x_i = x}(y) &= \sum_{t \in \mathcal [K]}\parens*{f_{\mathbf y_i| k, x}(y) - f_{\mathbf y_i| l, x}(y)}\prob{\mathbf t_i = t | \mathbf x_i = x}.
\end{align*}
Positivity gives that this quantity is well-defined for any $x \in \mathcal X$. Removing constants:
\begin{align*}
    f_{\mathbf y_i(k)|\mathbf x_i = x}(y) - f_{\mathbf y_i(l)|\mathbf x_i = x}(y) &= \parens*{f_{\mathbf y_i| k, x}(y) - f_{\mathbf y_i| l, x}(y)}\sum_{t \in \mathcal [K]}\prob{\mathbf t_i = t | \mathbf x_i = x} \\
    &= f_{\mathbf y_i| k, x}(y) - f_{\mathbf y_i| l, x}.(y),\,\,\,\,\sum_{t \in \mathcal [K]}\prob{\mathbf t_i = t | \mathbf x_i = x} = 1
\end{align*}
Therefore, $f_{\mathbf y_i(k)|\mathbf x_i = x}(y) \neq f_{\mathbf y_i(l)|\mathbf x_i = x}(y)$ for some $k, l \in [K]$ if and only if there exists some $x \in \mathcal X$ s.t. $f_{\mathbf y_i | k, x}(y) \neq f_{\mathbf y_i | l, x}(y)$.
\end{proofE}

\subsection{Hypothesis Testing}

By Lemma \ref{thm:CoDiCE}, under causal assumptions, conditional $K$-sample tests can be used for inference about potential outcomes. We can tie this into ith $K$-sample causal discrepancy testing, with \textit{any} consistent conditional independence test via the following corollary:
\begin{corollaryE}[Consistent Conditional Independence Tests and Consistent $K$-sample causal discrepancy tests]
Assume the setup described in Setup \ref{setup}. Further, suppose that:
\begin{enumerate}
    \item The treatment assignment is ignorable: $\parens*{\mathbf y_i(1), ..., \mathbf y_i(K)} \indep \mathbf t_i \cond \mathbf x_i$, and
    \item The treatment assignments are positive for all levels of the covariates: $\prob{\mathbf t_i = k \cond \mathbf x_i = x} > 0$ for any $x \in \mathcal X$.
\end{enumerate}
Then a consistent conditional independence test of Equation \eqref{eqn:hypo_cond_ind} is equivalent to a consistent $K$-sample causal conditional discrepancy test in Definition \ref{def:hypo_causal_kst}.
\label{cor:cond_ind}
\end{corollaryE}
\begin{proofE}
Note that with $\vec{\mathbf v}_i = g(\mathbf t_i)$ as-defined in Remark \ref{rem:ind_test_kst}, then direct application of Theorem \ref{thm:CoDiCE} followed by Remark \ref{rem:ind_test_kst} give that for some $k, l \in [K]$ and $x \in \mathcal X$, then $F_{\mathbf y_i(k)|\mathbf x_i = x} \neq F_{\mathbf y_i(l)|\mathbf x_i = x} \iff F_{\mathbf y_i | k, x} \neq F_{\mathbf y_i | l, x} \iff F_{\mathbf y_i, \mathbf v_i | \mathbf x_i} \neq F_{\mathbf y_i | \mathbf x_i} F_{\mathbf v_i | \mathbf x_i}$, as-desired.
\end{proofE}

Briefly, as long as we can identify a consistent conditional independence test for a related problem of identifying the independence of $\mathbf y_i$ and $\mathbf v_i$ conditional on the covariates $\mathbf x_i$, and the causal assumptions of ignorability and positivity are applicable, then we can obtain evidence for a $K$-sample causal conditional discrepancy on the basis of a conditional independence test. Note that this theorem provides guarantees for the hypotheses to be equivalent; it does \textit{not} make guarantees about a particular conditional independence test being consistent in a given setting. For instance, we may need further qualifications for a conditional independence test to be a consistent test, such as finite first and second moments of the outcomes and the baseline covariates.

\subsection{$K$-sample Causal Unconditional Discrepencies}
\label{sec:causal_uncond_discrep}
The above approaches readily generalize to discrepencies, in which $F_{\mathbf y_i(k)} \neq F_{\mathbf y_i(l)}$ for some $k, l$ with additional assumptions (called an unconditional causal discrepancy). Appendix \ref{app:udice} derives the above results and sufficient additional assumptions for the approaches described herein to the unconditional case. Intuitively, our main result is that tests for $K$-sample conditional discrepancy tests are equivalent to $K$-sample (unconditional) causal discrepency tests if differences in the covariate distributions (across groups) can be characterized by shifts in the density of the outcome (conditional on the covariates) that are in the same ``direction'' across all levels of the covariate(s) $\mathbf x_i$. 

Somewhat contrary to intuition, under standard causal assumptions, unconditional conditional discrepancy tests are \textit{not} equivalent to unconditional causal discrepancy tests without the restrictive assumption that these ``shifts'' are constant across all covariates. Under this more restrictive framework, a similar approach to corollary \ref{cor:cond_ind} gives that unconditional discrepancy tests can be used to test for unconditional causal discrepancies. In the more general case, as long as the discrepancy is in the same direction (and may be of different magnitudes), conditional discrepancy tests can be used for unconditional causal discrepancy tests. Together, these results give us the ability to characterize unconditional causal discrepancies to causal conditional discrepancies in much the same way as characterizations of the ATE to CATEs.

\section{Numerical Experiments}
\subsection{Conditional independence tests}
\label{sec:statistics}

In this paper, we consider the effectiveness of \cmanova, \kcd, \rcit, \rcot, \gcm, and \cdcorr~for $K$-sample causal conditional discrepancy testing. We also benchmark these strategies against the distance correlation, \Dcorr, which facilitates a strategy for unconditional causal discrepancy testing (see Appendix \ref{app:udice}). The reason that we compare to \Dcorr~is that it is possible that by ignoring covariates entirely, we may see increased finite-sample testing power due to computational efficiency under certain contexts, despite a loss of testing validity or power under others. See Appendix \ref{app:methods} for details on the methods and statistical tests employed.

\subsection{Causal Assumptions in Observational Studies}
\label{sec:matching}

The results discussed in Lemma \ref{thm:CoDiCE} and Corollary \ref{cor:cond_ind}, which allow us to make causal conclusions on the basis of the outcome of a $K$-sample conditional discrepancy test or conditional independence test, apply only in the event that the causal assumptions hold: \textit{ignorability}, \textit{positivity}, and \textit{no interference}. Other than the no interference criterion, both the ignorability and positivity assumptions can be readily reasoned in randomized trials: if we randomly assign people to a treatment or control group (conditionally or unconditionally on baseline covariates, as long as we \textit{measure} the covariates we use to randomly assign treatment or controls), both the ignorability and positivity assumptions can be satisfied \cite{Oakes2013}. The \textit{no interference} criterion can be reasoned through via domain expertise by noting whether or not the treatment group of individuals impacts the outcomes of other individuals \cite{Rosenbaum2007Mar}. Consider, for instance, a case where we want to measure the impact of a vaccine, and we treat individuals with a vaccine who will only be exposed to other vaccinated individuals. The no interference assumption could be violated since the treatments of the vaccinated individuals could impact the potential outcomes of the other individuals.

However, what happens if our data is not randomized, but is observational, in that the researcher does not have explicit control over ensuring treatment assignments in a randomized fashion? Intuitively, ignorability can be conceptualized as \textit{no unmeasured confounding}, in that any confounding variables must be collected in the observed data \cite{Greenland2009}. This criterion can be limiting in that while strong assumptions can be made about unobserved variables, it is unverifiable in the obtained data sample. Causal inference in these observational settings is therefore limited by the sensitivity of the ignorability assumption to these unmeasured variables \cite{Rosenbaum1984Mar}. The no interference criterion is similar in interpretation as it is for the randomized trial case. Ignorability aside, the positivity assumption can prove difficult to reason through. Conceptually, positivity asserts that for all levels of the covariate under study, there is a non-zero probability of a sample obtaining the treatment \textit{or} the control. This assumption is impossible to verify in an observational study, since it is statistically a \textit{pre-hoc} criterion. However, we can take principled approaches to identify subsamples of the data upon which this might be the case. Attempts to rectify this assumption are practically addressed via matching \cite{Petersen2012Feb,Stuart2010Feb}.

Through matching, we attempt to identify a subset of the samples from an observational study upon which positivity \textit{might} hold by trimming subjects whose treatment group assignment is deterministic in the empirical sample based on their observed covariates \cite{Kang2016Jan}. A number of techniques have been proposed to ascertain positivity from observational studies \cite{Stuart2010Feb}. In this investigation, we choose to address this via vector matching (\vm) as proposed by \citet{Lopez2014}, which is a strategy for matching subjects across multiple treatments and is closely related to propensity score matching for more than two treatments. In the case where the treatment is nominal, the generalized propensity score $r(t, x)$ is the probability $Pr(\mathbf t = t | \mathbf x = x)$ of being assigned to treatment group $t$ given the baseline covariates $x$. For a given individual with baseline covariates $x_i$, a set of generalized propensity scores $\mathcal R(x_i) = \left\{\hat r(t, x_i)\right\}_{t \in [K]}$ are estimated using a multinomial regression model. For each treatment $t \in [K]$, we compute the following quantities:
\begin{align*}
    l(t) &\triangleq \max_{t' \in [K]}\parens*{\min_{i \in [n] : t_i = t'}\set*{\hat r(t, x_i)}},\,\,
    h(t) \triangleq \min_{t' \in [K]}\parens*{\max_{i \in [n] : t_i = t'}\set*{\hat r(t, x_i)}} \numberthis \label{eqn:vm}
\end{align*}
Individuals $i$ with $\hat r(t, x_i) \not \in \left(l(t), h(t)\right)$ for any $t \in [K]$ are discarded from successive hypothesis testing. Intuitively, \vm~ensures that no retained individuals have covariates which occur with extremely low probability (or extremely high probability) for any particular group. Strategies which first pre-process observational data using \vm~are henceforth referred to as ``causal'', and we use this strategy for \ccdcorr. The model employed for multinomial regression is given in Appendix \ref{app:sims}.

\subsection{Simulations}
\label{sec:sims}

\begin{figure}[h!]
    \centering
    \includegraphics[width=\linewidth]{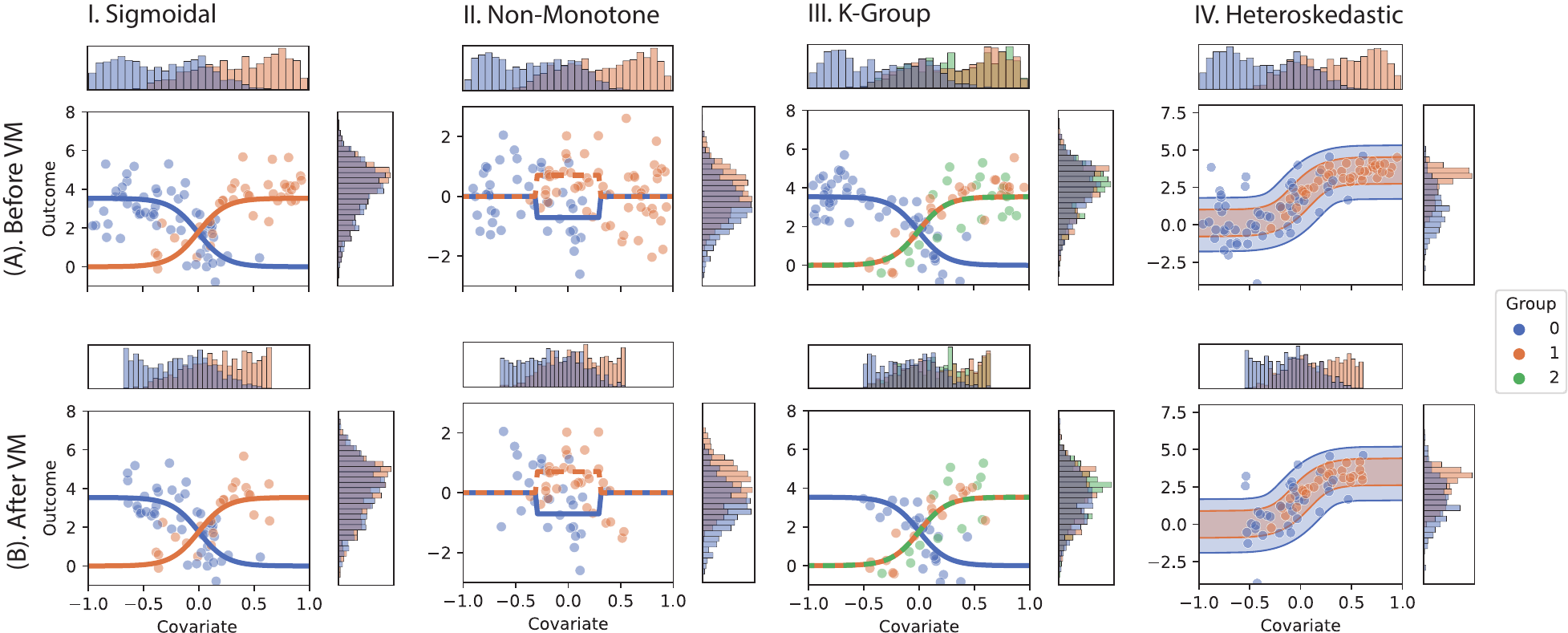}
    \caption{\textbf{Four simulations to empirically evaluate causal conditional discrepancy tests.} $n=100$ samples are selected from one of four simulation settings (\textbf{I.} to \textbf{IV.}) for a given degree of covariate balance, dimensionality, and effect size $\Delta$. Each point shows a single sample from the given model. For \textbf{I.}-\textbf{III.}, the solid line indicates the true outcome for a particular group at a given covariate level. For \textbf{IV.}, the opaque boxes indicate the standard deviation from the mean for each group. All simulations are shown for $\Delta = 1.0$ with low balance (40\% covariate distribution overlap). Appendix \ref{app:sims} illustrates the simulations for ranges of effect size from $0.0$ to $1.0$ to demonstrate the causal conditional discrepancy that should be detected. Marginal histograms are indicated in both the outcome and the covariate dimension. \textbf{(A)} illustrates all $n=100$ samples before \vm, and \textbf{(B)} illustrates the samples retained after \vm. \vm~trims the samples to have alignment (across groups) with respect to the covariate distribution.}
    \label{fig:simsetup}
\end{figure}

We empirically investigate the flexibility, validity, and accuracy of $K$-sample causal conditional discrepancy testing using simulations that extend beyond our theoretical claims and mirror observational frameworks. In Figure \ref{fig:simsetup}, we illustrate the simulations under which our proposed techniques are evaluated. $n=100$ samples are collected from the indicated statistical model, with variable ``dimensionality'' and ``balance''. The ``dimensionality'' indicates the number of dimensions of the outcome, and the ``balance'' indicates the level of similarity for the covariate distributions of the different treatment groups. The $y$-axis indicates the outcome in the first dimension, and the $x$-axis indicates the covariate associated with a single sample point. The solid line indicates the average outcome for a given group at a particular covariate level. The causal conditional discrepancy that we wish to detect is illustrated by the difference between these two lines for a given covariate level (the \textit{signal}).

For successive dimensions, the relationship between the outcome and the covariate for a given group remains the same; however, the signal to noise ratio (per-dimension) decreases as the dimensionality increases. We run each simulation in a low dimensionality ($D = 10$) and high dimensionality ($D = 101$) regime. The dimensionality for high dimensional simulations was chosen such that the simulation exhibits the high-dimensionality, low sample size phenomenon (HDLSS) \cite{Hall2005}, which presents a challenge for parametric techniques as the number of dimensions exceeds the number of samples. Hence, \cmanova-like strategies cannot be employed without restrictive assumptions on the true underlying model \cite{Chi2013}, which may not be practical for many high-dimensional datasets (such as connectomics or genomics datasets) or new datasets which are not yet fully understood. 

Figure \ref{fig:simsetup}\textbf{.I.} \textit{Sigmoidal} indicates a simulation with a sigmoidal relationship between the outcome at a given dimension and the covariate. The effect size measures the degree to which the second group is \textit{rotated} about the other group (effect size of $0$ corresponds to the two distributions being identical, and effect size of $1$ corresponds to the second group being rotated $180$ degrees). This simulation was designed to test the sensitivity of the included tests to a causal conditional discrepancy when there is no unconditional causal discrepancy, as effect sizes of both $0$ and $1$ correspond to the unconditional outcome distributions being identical across both groups across all dimensions. This is due to the fact that the effect that is introduced is a rotation of the outcomes about a particular covariate level. Figure \ref{fig:simsetup}\textbf{.II.} \textit{Non-Monotone} shows a simulation with a non-monotone relationship between the outcome at a given dimension and the covariate. In this case, an unconditional causal discrepancy exists. We would expect that high performing techniques will perform as well as or better than unconditional causal discrepancy tests in this context. Figure \ref{fig:simsetup}\textbf{.III.} \textit{K-Group} indicates a simulation with three groups, wherein the first group has a different covariate distribution from the second two groups. The effect size again measures the degree to which the first group is rotated about the other two groups (as in \ref{fig:simsetup}\textbf{.I.} Sigmoidal). Figure \ref{fig:simsetup}\textbf{.IV.} \textit{Heteroskedastic} indicates a simulation where the covariance of the outcome for one group exceeds that of the other group for each possible value of the covariate, despite the fact that the average outcome (conditional on the covariate value) is identical across both groups. The effect size measures how many times larger the covariance is for the first group than the second group. 

These simulations were conducted under various balance contexts, where the balance indicates the fraction of samples which have the same covariate distribution. The remaining samples are collected from asymmetric covariate distributions. When balance is low (shown in Figure \ref{fig:simsetup}(A), with a balance of $40\%$) this induces group-specific imbalance in which the treatment groups being compared do not have common support. For \ccdcorr, the samples are first filtered using \vm, as described in Methods \ref{sec:matching}. The effects of this pre-processing step are indicated in Figure \ref{fig:simsetup}(B). Note that \vm~pre-processes the samples so that both groups have common covariate support.

The outcomes are finally randomly rotated in $D$-dimensional space to ensure that the techniques are incorporating information across dimensions (as the top dimension contains more signal than successive dimensions) \cite{haar}. Appendix \ref{app:sims} provides technical details for each simulation employed, and illustrates similar plots across a range of effect sizes for a given setting.

\section{Results of Numerical Experiments}

\subsection{\ccdcorr~provides a substantial improvement in finite-sample validity over \cdcorr}

In Corollary \ref{cor:cond_ind}, we made the observation that if ignorability and positivity hold along with the assumptions of Setup \ref{setup}, then a consistent conditional independence test is a test for a $k$-sample causal discrepancy test. Conceptually, what this means is that as the sample size grows, the $k$-sample causal conditional discrepancy test is informative for differentiating whether or not the data provides sufficient information to reject the null hypothesis of interest in Definition \ref{def:hypo_causal_kst} for an acceptable type I error threshold $\alpha$. In an observational dataset, however, we simply observe the data: we do not have any information as to whether the positivity criterion is satisfied, and we do know whether sample size is sufficient for our causal conditional discrepancy test to be informative. Therefore, the ability to conduct valid inference in the absence of omniscent knowledge about the positivity condition and for datasets with a finite number of samples is imperative. We turn to simulation to investigate the implications of a lack of covariate balance on the empirical validity. To investigate the empirical validity, we run all simulations with an effect size $\Delta = 0$ (the outcome distributions conditional on the covariates are identical, and $H_0$ is true) and decrease the covariate balance from $1$ (the covariate distributions are identical across \textit{all} groups) to $0.2$ (the covariate distribution differs for one of the groups for all but $20\%$ of samples). Under this framework, empirically valid tests for causal conditional discrepancies will reject the null hypothesis in favor of the alternative at a rate $\leq \alpha$ irrespective of the level of covariate balance (the \textit{Type I Error Rate} for a given simulation at a given level of covariate balance). Conceptually, when covariate balance is low, it may not be possible to ascertain whether differences between the groups are due to the covariate distribution asymmetries \textit{or} the group assignment due to confounding by the covariate. Tests which make a type I error at a rate $>\alpha$ may be \textit{aliasing} the outcome/group effect (conditional on the covariate) with the outcome/covariate effect.

\begin{figure}
    \centering
    \includegraphics[width=\linewidth]{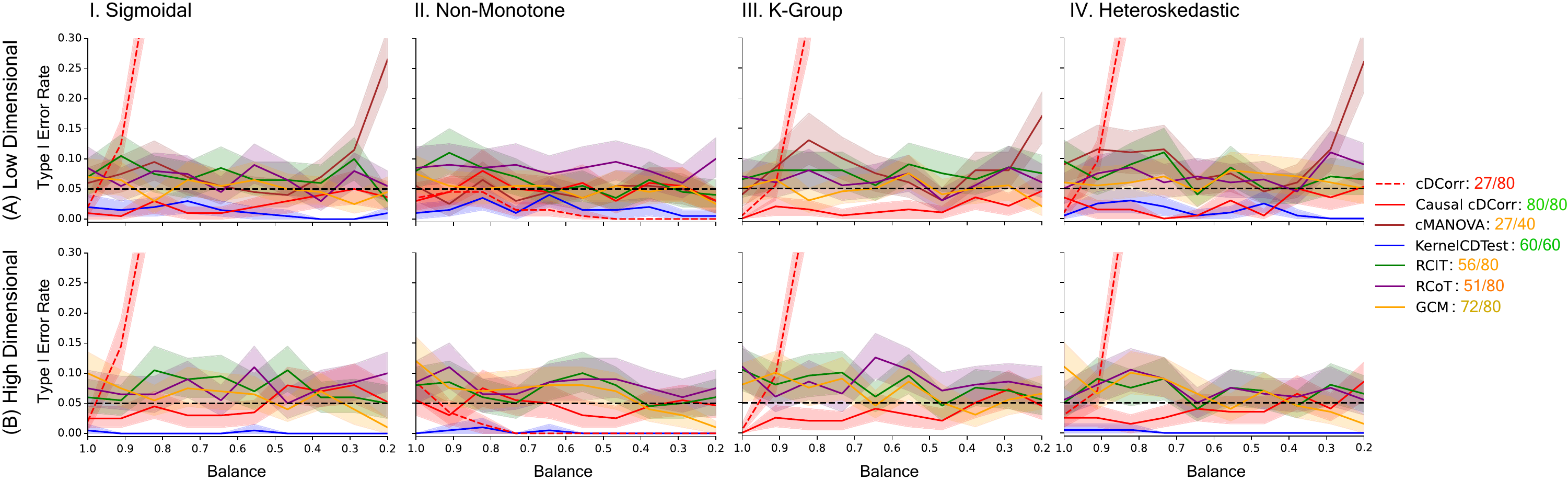}
    \caption{\textbf{\ccdcorr~and \kcd~provide valid statistical tests across multiple simulation settings for $k$-sample causal conditional discrepancy testing}. The four simulation settings in \textbf{I.} to \textbf{IV.} in low \textbf{(A)} and high-dimensional \textbf{(B)} contexts. Valid statistical tests will tend to reject the null hypothesis in favor of the alternative at a rate $\leq \alpha$ when no effect is present. Each line reflects the type I error rate of a particular test (line color) at a given balance for $\alpha = 0.05$ (black dashed line) when the effect size $\Delta = 0$. The ribbon indicates a 90\% confidence interval of the estimated type I error rate estimate over $R=100$ repetitions. The legend at the right indicates the line color for the statistical test, as well as the fraction of times that the lower limit of the 90\% CI falls $\leq \alpha$ across (simulation, dimensionality, balance) settings. When balance is high, \cdcorr~is always valid, but as the balance decreases, \cdcorr~is almost never valid (with the exception of the ``non-monotone'' setting, \textbf{II.}). Likewise, \cmanova, \rcot, \rcit, and \gcm~do not achieve empirical validity in at least $75\%$ of all settings. Only \ccdcorr~and \kcd~show reasonably high empirical validity ($>75\%$) for causal conditional discrepancy testing.}
    \label{fig:validity}
\end{figure}

In Figure \ref{fig:validity}, we explore the statistical validity of all proposed conditional independence tests in various $k$-sample causal discrepancy testing settings by examining the type I error rate with the effect size $\Delta = 0$ (\textit{no effect} is present). Empirically valid statistical tests will make a type I error at a rate $\leq \alpha = 0.05$. We decrease the balance between the groups from $1.0$ to $0.2$, and compute estimates of the type I error rate. For each statistic, we also assess the frequency of type I errors by computing the number of times that the lower limit of a 90\% confidence interval (Wald test) is not $\leq \alpha = 0.05$. As the balance between the different groups decreases, \cdcorr~tends to make type I errors at an increasing rate (and, indeed, although the plots have been ``cut-off'' at a type I error rate of $0.3$ in favor of increased resolution near $\alpha = 0.05$ for the other techniques, the ``cut-off'' lines all reach a power of $1.0$ at or before the balance reaches its minimum value of $0.2$). \cmanova, \rcit, and \rcot~ frequently make type I errors at a rate $> \alpha$ over $25\%$ of the time. \gcm~makes type I errors infrequently (about 10\% of the time). By augmenting \cdcorr~with \vm~to produce \ccdcorr, the validity improves markedly: \cdcorr~goes from frequently making type I errors (53 of 80 possible settings) to never making type I errors (0 of 80 settings). \kcd~also never makes type I errors, but is not amenable to $K$-group settings.

\subsection{Causal \cdcorr~ empirically dominates other techniques for causal conditional discrepancy testing}

These simulations were conducted under two balance contexts (high and low balance) in which the covariate distribution is equal for a high portion ($80\%$) and a low portion ($40\%$) of the samples, with the remaining samples collected from asymmetric covariate distributions. In the low balance case (shown in Figure \ref{fig:simsetup}A) this induces group-specific imbalance in which the treatment groups being compared do not have common support. For Causal~strategies, the samples are first filtered using \vm, as described in Methods \ref{sec:matching}. The effects of this pre-processing step are indicated in Figure \ref{fig:simsetup}B.

\begin{figure}[h!]
    \centering
    \includegraphics[width=\linewidth]{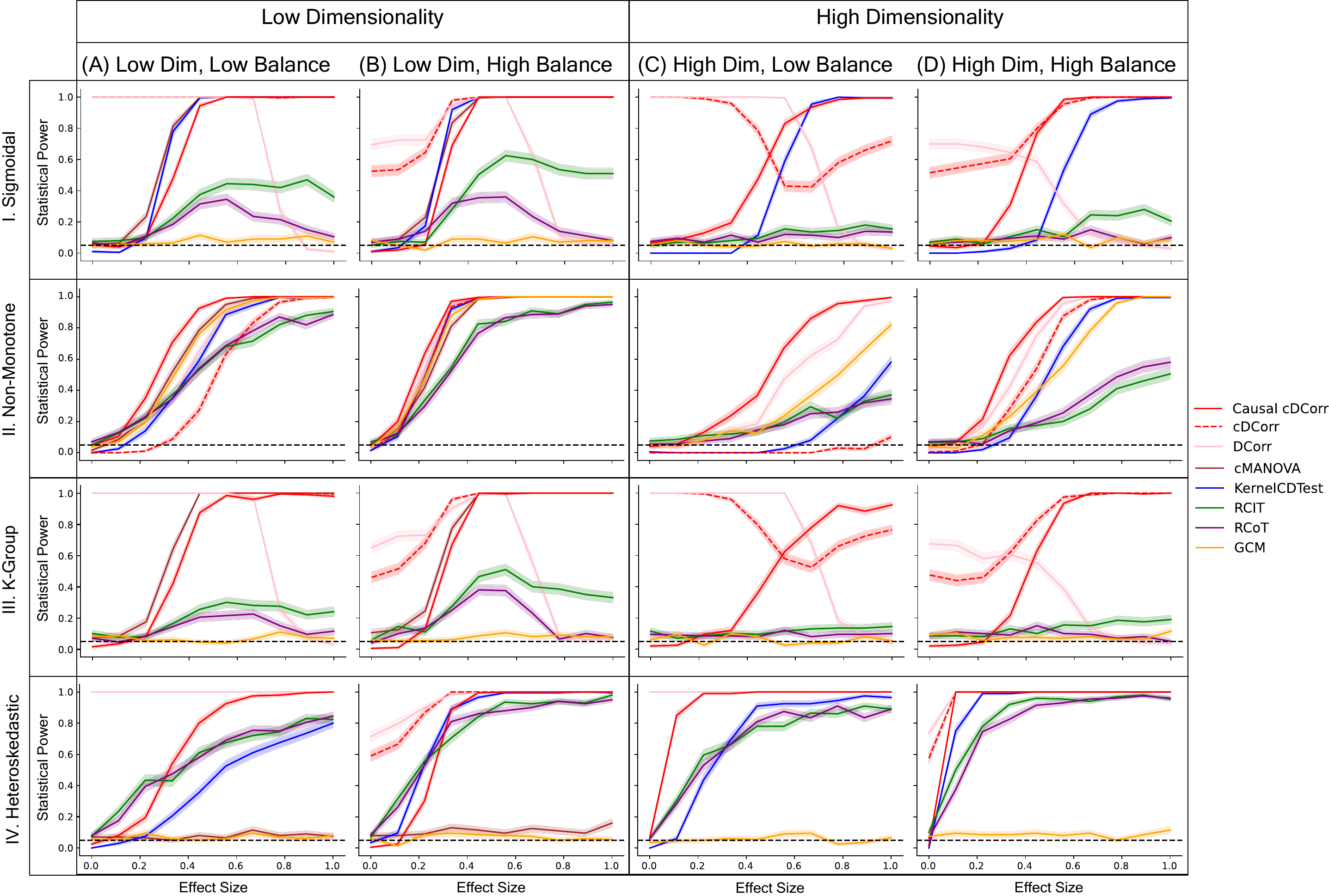}
    \caption{\textbf{\ccdcorr~provides a flexible statistical test for causal effect detection which balances statistical power and validity}. $n=100$ simulations are sampled from four regimes, under two levels of covariate overlap (low and high covariate balance), and two levels of dimensionality (low and high dimensionality), each for $R=200$ repetitions. The effect size $\Delta$ ($x$-axis) for a particular simulation setting indicates the disparity between the different groups, and the statistical power for a given technique (colored lines) indicates the fraction of repetitions in which the test indicated a $p$-value $\leq \alpha$ ($y$-axis). $\alpha = 0.05$ is indicated with the black dashed line. Colored ribbons reflect a 90\% confidence interval of the estimated statistical power. $\Delta = 0$ corresponds to the groups being identical (valid tests will reject the null hypothesis at a rate $\leq \alpha$). As $\Delta$ increases to $1$, powerful tests will see the statistical power increase. \kcd~appears to be reasonably powerful, but does not operate on the $k$-group simulation. \ccdcorr~is the only powerful test in all simulations.}
    \label{fig:powers}
\end{figure}

For a given effect size and simulation setting, statistical power is estimated over $R=200$ repetitions at $\alpha = 0.05$ and is shown in Figure \ref{fig:powers}. Validity (Effect size $\Delta = 0.0$) is investigated thoroughly in Figure \ref{fig:validity}. \cdcorr~and \Dcorr~(which can be used as tests for unconditional causal discrepancies, rather than causal conditional discrepancies; Appendix \ref{app:udice}) are invalid in numerous contexts, such as Figure \ref{fig:powers}\textbf{.I.} and Figure \ref{fig:powers}\textbf{.III.}, the sigmoidal and $K$-group settings. In these cases, type I error can be ascertained from the statistical power with an effect size of $\Delta = 0$. In particular, note that the \Dcorr~power curves appear particularly puzzling: the power is maximal at $\Delta = 0.5$, and falls for larger or smaller effect sizes. These simulations were constructed such that at $\Delta = 1.0$, the \textit{unconditional} causal discrepancy was minimized, despite the fact that the distributions are conditionally (on the covariates) maximally dissimilar. At $\Delta = 0.0$, an \textit{unconditional} effect appears to be present at a rate $\geq \alpha$ simply because the group conditional covariate distributions are different. This can be conceptualized as the unconditional effect (between the outcome and group) being \textit{aliased} with the outcome/covariate effect due to the disparate covariate distributions. Appendix \ref{app:sims} provides further clarity as to \textit{why} the the unconditional distributions are identical across all groups when $\Delta = 1.0$ for the sigmoidal and $K$-group settings. Causal \cdcorr, both variations of \cmanova, and \kcd~are the only approaches which are valid under all contexts. This result provides significant caution to the use of unconditional effect tests in data with conditional effects, as this unanticipated ``effect aliasing'' may arise.

Of the remaining tests, \gcm~has almost no power in any of the simulations except for the non-monotone simulation. \rcit~and \rcot~have almost no power in the $k$-group settings, and have low power in the sigmoidal and non-monotone setting when dimensionality is high. \kcd~tends to be overly conservative in contexts in which an effect is present, and generally has much lower power than Causal \cdcorr~(As shown in Figure \ref{fig:powers}\textbf{.I.}, Figure \ref{fig:powers}\textbf{.II.}, and Figure \ref{fig:powers}\textbf{.IV.}) across all but Figure \ref{fig:powers}\textbf{.I.(A)}, Figure \ref{fig:powers}\textbf{.I.(B)}, and Figure \ref{fig:powers}\textbf{.IV.(B)} where their disparity is relatively minor. In particular, Causal \cdcorr~is \textit{always} more powerful in high dimensional regimes of Figure \ref{fig:powers}\textbf{.(C)-(D)}. Further, Causal \cdcorr~flexibly operates in the $K$-group regime Figure \ref{fig:powers}\textbf{.III.}, for which there is no analogous extension of \kcd. \cmanova~approaches have no power in regimes in which the causal conditional discrepancy can be categorized as a difference in higher order moments, as in \textbf{IV.}, in which the two groups differ only in their second moment. Finally, \cmanova~approaches cannot be natively extended to the HDLSS regimes of Figure \ref{fig:powers}\textbf{.(C)-(D)} without further considerations, as-noted in Section \ref{sec:statistics}. Taken together with the results of Figure \ref{fig:validity}, this suggests that augmentation of \cdcorr~with \vm~to produce \ccdcorr~provides a flexible, valid, and powerful test for uncovering $k$-sample causal conditional discrepancies across a range of contexts, including when the covariate/outcome relationship is non-linear or non-monotone, and when the relationship across groups manifests in higher-order moments.

%

\section{Discussion}

In this work, we introduce a conceptual framework for defining and understanding causal sources of heterogeneity in outcomes across nominal treatment regimes. This work builds upon the efforts of \citet{Park2021Jul}, and provides a theoretical framework in which conditional independence tests produce $k$-sample causal conditional discrepancy tests. We give sufficient conditions for $k$-sample (unconditional) causal discrepancy tests to be equivalent to $k$-sample causal conditional discrepancy tests, which generalizes the relationship between the ATE and the CATE to multivariate frameworks.  While the theoretical framework which we elucidated is almost certainly violated in observational investigations, we demonstrate how augmentation of conditional independence tests via \vm~yields tests which achieve both high power and empirical validity in finite sample contexts. These features are often missed entirely by other methods; in particular, in a number of contexts power decreases as the effect size increases, indicating that they are extremely subject to biases due to confounding. In this sense, while they may be theoretically powerful when $n$ is arbitrarily large, they are extremely subject to finite-sample issues when confounding is present. Together, we believe that this work demonstrates the utility of devising assumption-driven statistical tests for conditional independence testing on the basis of causality.

\paragraph*{Connections to causal structure learning and causal discovery}

The framework that we have introduced features direct applications in the field of constraint-based causal discovery. Typically, constraint-based causal discovery is implemented via incorporation of conditional independence tests with strategies such as the PC or the FCI algorithms \cite{Spirtes,Spirtes2001Jan}. 

These procedures typically decompose the sets of variables under study into nodes in a graph, and proceed to iteratively build a graph of causal dependencies (a directed acyclic graph, or \texttt{DAG} \cite{Pearl2009Jan,Pearl2010Jul}) through progressively more restrictive conditional independence tests (e.g., iteratively ``building'' a proper conditioning set $\mathbf x_i$ from the observed data through sequences of conditional independence tests). In the case where sets of variables that we wish to test are nominal, the conditional independence tests used are conditional discrepancy tests, of which we have directly compared our techniques to, such as \kcd~\cite{Park2021Jul}. Through elucidating explicit statistical assumptions necessary for conditional independence tests to be causal conditional discrepancy tests, we were able to yield a new test, \ccdcorr, by attempting to faithfully modify the observed data to better reflect the underlying statistical assumptions. \ccdcorr~showed substantial improvements in both finite sample sensitivity and specificity (it shows higher validity in null experiments, and power generally increases with effect size), and generalizability (it applies readily to greater than $2$-level contexts, where we have more than binary treatment levels), while maintaining empirical validity. 

Inherently, the success or failure at recovering the underlying causal structure is directly a function of the sensitivity and specificity of the conditional independence tests leveraged to true underlying dependencies in the data. We believe that this demonstrates that our assumption-driven statistical framework may yield fruitful improvements to causal structure learning as it relates to nominal variables, and will allow our methodologies to inform new developments in this exciting avenue of work.

\paragraph*{Limitations and future work}

A chief limitation of this work is that we largely focus on the nominal treatment case; e.g., where $\mathbf t_i$ is a $[T]$-valued random variable. Indeed, it is often the case where one may wish to investigate continuous or multivariate treatments through more general conditional independence tests. In the case of continuous or multivariate $\mathbf t_i$, the positivity assumption becomes one of densities rather than probabilities. With $(\mathcal T, \delta_t)$ a metric space and $\mathbf t_i$ a $\mathcal T$-valued random variable, the positivity assumption states $f_{\mathbf t_i | x}(t) > 0$ for all $t \in \mathcal T$ and for all $x \in \mathcal X$. In light of Lemma \ref{thm:CoDiCE}, our sums become integrals, and the results that we have developed herein remain otherwise identical. 

Due to the flexibility of \cdcorr, the approach upon which \ccdcorr~was based, we believe that this limitation can be overcome with similar pre-processing and analytical strategies to how we worked towards better reflecting the positivity assumption in observational data in the nominal case. Many strategies exist which attempt to bin continuous or multivariate treatments into discrete bins (either through stratification or binning of the treatment variable in the continuous or multivariate case), which would directly extend under the framework which we have proposed. However, the utility of these strategies are dubious due to their tendency to impart strong biases and a resulting loss of statistical power \cite{Royston2006Jan}. Both parametric and non-parametric strategies exist which broadly attempt to ``re-weight'' the observed samples in observational studies which can be empirically conceptualized as assumption-driven approaches working towards the positivity assumption from observational data. These similarly leverage the generalized propensity score, where $r(t, x) = f_{\mathbf t_i | x}(t)$ \cite{Robins1986Jan,Imai2004Sep,Hirano2004Jul} is a density instead of a probability. These techniques include entropy balancing \cite{Vegetabile2021Mar,Tubbicke2020Oct} and more classical weight-stabilization approaches \cite{Robins2000Sep}. Natively, the \cdcorr~procedure \cite{Wang2015} directly incorporates weight to reflect conditional structures in the model. The non-parametric models used to infer weights could be directly altered to reflect better finite-sample characteristics, and thereby directly tune for positivity, using the generalized propensity score while maintaining the asymptotic theoretical guarantees afforded by the traditional \cdcorr~approach. 

\paragraph{Acknowledgements}

The authors are grateful for the support from the National Science Foundation (NSF) administered through NSF Career Award NSF 17-537, the National Institute of Health (NIH) through National Institute of Mental Health (NIMH) Research Projects 1R01MH120482-01 and RF1MH128696-01, and the NIH through Research Project RO1AG066184-01. 

\paragraph{Code and Data Availability Statement}

All figures within this manuscript can be reproduced via the github repository at \href{https://github.com/ebridge2/cdcorr/}{github.com/ebridge2/cdcorr/}. 

\paragraph{Author Contributions}

EWB wrote the paper; EWB and JTV revised the manuscript; EWB and JTV conducted study conception; EWB conducted study design; EWB interpreted the results; EWB, JC, and SP contributed code; EWB and JTV devised the statistical methods; AB processed and interpreted the mouse connectomes; EWB analyzed the data; BG, AL, CS, BC, and JTV provided valuable feedback and discussion throughout the work.

\bibliographystyle{IEEEtranSN}
\bibliography{batch}

\newpage

\appendix
\section{Theoretical Results}

Under the restriction that $\mathbf y_i(2)$ and $\mathbf y_i(1)$ differ at most by an offset, the hypothesis in Equation \eqref{eqn:hypo_causal_2s} is exactly equivalent to a test of whether there is a CATE:

\begin{remark}
Suppose that if $\mathbf y_i(1) | \mathbf x_i = x \distas{d} F_{\mathbf y_i(1) | \mathbf x_i = x}$, then there exists a constant $\epsilon_x$ s.t. if $\mathbf x_i = x$, then $\mathbf y_i(2) = \mathbf y_i(1) + \epsilon_x$. Then $\gamma_x \neq 0$ if and only if $F_{\mathbf y_i(1) | \mathbf x_i = x} \neq F_{\mathbf y_i(2) | \mathbf x_i = x}$.
\end{remark}
\begin{proof}

$\Rightarrow$) Suppose that $\gamma_x \neq 0$.

Then $F_{\mathbf y_i(1) | \mathbf x_i = x} \neq F_{\mathbf y_i(2) | \mathbf x_i = x}$, since two random variables differing in expectation implies a difference in distribution.

$\Leftarrow$) Suppose that $F_{\mathbf y_i(1) | \mathbf x_i = x} \neq F_{\mathbf y_i(2) | \mathbf x_i = x}$, and further for any $\mathbf y_i(1) | \mathbf x_i = x \distas{d} F_{\mathbf y_i(1) | \mathbf x_i = x}$, that if given $\mathbf x_i = x$, $\mathbf y_i(2) = \mathbf y_i(1) + \epsilon_x$ for some $\epsilon_x$.

Since $F_{\mathbf y_i(1) | \mathbf x_i = x} \neq F_{\mathbf y_i(2) | \mathbf x_i = x}$, then $\epsilon \neq 0$, as otherwise $F_{\mathbf y_i(2) | \mathbf x_i = x} = F_{\mathbf y_i(1) | \mathbf x_i = x}$.

Then:
\begin{align*}
    \gamma_x &= \expect{\mathbf y_i(2) - \mathbf y_i(1) \cond \mathbf x_i = x} \\
    &= \expect{\epsilon_x} = \epsilon_x \neq 0,
\end{align*}
as desired.
\end{proof}
Note that the forward direction of this proof does not require any of the assumptions that $\mathbf y_i(2)$ is an offset of $\mathbf y_i(1)$, which demonstrates that the $2$-sample test shown in Equation \eqref{eqn:hypo_causal_2s} is a less-restrictive characterization than a CATE in \eqref{eqn:hypo_cate}.

\printProofs

\section{Methods}
\label{app:methods}
\subsection{\Dcorr} The distance correlation \cite{Szekely2007Dec} is implemented using the \sct{hyppo}~package \cite{Panda2019Jul}. $p$-values are estimated using \cite{Shen2022Jan}.

\subsection{\cmanova}
\label{app:methods:cmanova} With the hypothesis in Definition \ref{def:hypo_causal_kst} in mind, we  investigate whether an ``alternative model'' including, for each dimension in the dataset, an intercept, a slope associate with the covariates, a group-specific intercept, and an interaction term between the group and the covariate is significant against a ``null model'' which includes only an intercept and a slope associated with the covariates. Conceptually, this corresponds to investigating whether the two groups differ (either conditionally or unconditionally on the covariates). 

With $\Sigma_{A,m}$ and $\Sigma_{0,m}$ the product of model variance for alternative and the null models respectively, under standard assumptions of multivariate least-squares regression (which are violated by the below-described simulations), the test in Definition \ref{def:hypo_causal_kst} is equivalent to testing:
\begin{align*}
    H_0 : \Sigma_{A,m} = \Sigma_{0,m}\text{ against }H_A : \Sigma_{A,m} \neq \Sigma_{A,m}. \numberthis \label{eqn:manova_hypo}
\end{align*}
Denote $\Sigma_{A, E}$ and $\Sigma_{0,E}$ to be the error of the variance matrix under the alternative model and the null model respectively, and let $B_j = \frac{\Sigma_{j,m}}{\Sigma_{j,E}}$. Note that if $D$ is the number of dimensions of the outcome, the error of the variance matrix is a $D \times D$ matrix, and will be non-invertible if $D > n - m_j$, where $n$ is the number of samples and $m_j$ is the number of parameters in model $j$. In the simulations described, this occurs for the high dimensional case (where $D = 100 = n$), so \cmanova~cannot be used. Under this framework, given that the models are nested, the Pillai-Bartlett trace \cite{Pillai1959,Pillai1976,Pillai1977} is defined:
\begin{align*}
    \Lambda_{Pillai}^{A,0} &= \Lambda_{Pillai}^A - \Lambda_{Pillai}^0 \\
    &= \trace{B_A\parens*{I_D + B_A}^{-1}} - \trace{B_0\parens*{I_D + B_0}^{-1}}
\end{align*}
The statistic $\Lambda_{Pillai}$ can be used to test the hypothesis in Equation \eqref{eqn:manova_hypo}
The statistic $\Lambda_{Pillai}$ can be used to test the hypothesis in Equation \eqref{eqn:manova_hypo} by rescaling to a statistic which is approximately $F$-distributed \cite{Pillai1959,Pillai1976,Pillai1977}. \cmanova~in this manuscript is performed using the \sct{anova.mlmlist} (analysis of variance for nested multivariate linear models) function in \sct{R}, provided by the \sct{stats} package \cite{Fox2018Oct,stats}.

\subsection{\rcit~and \rcot} \rcit~and \rcot~are implemented using the \sct{RCIT}~package \cite{Strobl2019Mar} in \sct{R}. $p$-values for both techniques are estimated using a permutation test, with the number of null replicates $N=1000$. All other settings leverage the default values.

\subsection{\gcm} \gcm~is implemented using the \sct{GeneralisedCovarianceMeasure}~package \cite{Shah2018Apr,Peters2022}. The regression method leveraged is ``xgboost'' \cite{xgboost}, and $p$-values are estimated using a permutation test, with the number of null replicates $N=1000$. All other settings leverage the default values.

\subsection{\kcd} \kcd~\cite{Park2021Jul} is implemented using the \sct{doDiscover}~package \cite{dodisc}. $p$-values are estimated using a permutation test, with the number of null replicates $N=1000$. All other settings leverage the default values.

\subsection{\cdcorr} \cdcorr~\cite{Wang2015} is implemented using the \sct{hyppo}~package \cite{Panda2019Jul}. The distance metric used is the Euclidean distance, and $p$-values are estimated using a permutation test, with the number of null replicates $N=1000$. All other settings leverage the default values.

\section{Unconditional Discrepancies}
\label{app:udice}
For the unconditional case, we assume the setup described in \ref{setup}.

\begin{definition}[Unconditional $k$-sample causal discrepancy]
Suppose the setup described in \ref{setup}. An unconditional causal discrepancy exists if for any $k, l \in [K]$:
\begin{align*}
    F_{\mathbf y_i(k)} \neq F_{\mathbf y_i(l)}.
\end{align*}
\label{def:udice}
\end{definition}
and is hereafter referred to as a $k$-sample causal discrepancy (with no qualifiers about conditionality). 

This implies a natural hypothesis test for a causal discrepancy of:
\begin{align}
    H_0 : F_{\mathbf y_i(k)} = F_{\mathbf y_i(l)} \text{ for all $k, l$}\text{   against   }H_A : F_{\mathbf y_i(k)} \neq F_{\mathbf y_i(l)}\text{ for some }k, l\label{eqn:hypo_udice}
\end{align}

Consistent tests of Equation \eqref{eqn:hypo_udice} and Definition \ref{def:hypo_cond_kst} are equivalent if further causal assumptions are satisfied:
\begin{lemma}[Consistent $k$-sample conditional discrepancy tests and $k$-sample causal discrepancies]
\label{lem:udice}
Assume the setup described in \ref{setup}. Further, suppose that:
\begin{enumerate}
    \item The treatment assignment is ignorable: $\parens*{\mathbf y_i(1), ..., \mathbf y_i(K)} \indep \mathbf t_i \cond \mathbf x_i$,
    \item The treatment assignments are positive for all levels of the covariates: $\prob{\mathbf t_i = k \cond \mathbf x_i = x} > 0$ for any $x \in \mathcal X$,
    \item The conditional effect of treatment is equivalent in \textit{direction} across all covariate levels; e.g., for a given $y \in \mathcal Y$, for all $x \in \mathcal X$:
    \begin{align*}
        f_{\mathbf y_i | k, x}(y) - f_{\mathbf y_i | l, x}(y) \geq 0 \text{   or   }f_{\mathbf y_i|l, x}(y) - f_{\mathbf y_i | k, x}(y) \geq 0,
    \end{align*}
    and the inequality is strict for some $x \in \mathcal X$ where $f_{\mathbf x_i}(x) > 0$.
\end{enumerate}
Then a consistent test of Equation \eqref{eqn:cond_kst_discrep} for a conditional discrepancy is equivalent to a consistent test of Equation \eqref{eqn:hypo_udice} for an unconditional causal discrepancy.
\end{lemma}
\begin{proof}
By definition, $f_{\mathbf y_i(k)}(y)$ can be expressed as a marginalization over the joint density $f_{\mathbf y_i(k), \mathbf x_i, \mathbf t_i}(y,t,x)$ with respect to $\mathbf x_i$ and $\mathbf t_i$:
\begin{align*}
    f_{\mathbf y_i(k)}(y) &= \int_{\mathcal X \times \mathcal T}
    f_{\mathbf y_i(k) | t, x}(y)f_{\mathbf x_i, \mathbf t_i}(x, t)\,\text d (x,t),
\end{align*}

Using the definition of conditional probability gives:
\begin{align*}
    f_{\mathbf y_i(k)}(y) &= \int_{\mathcal X \times \mathcal T}
    f_{\mathbf y_i(k) | t, x}(y)\prob{\mathbf t_i = t | \mathbf x_i = x}f_{\mathbf x_i}(x)\,\text d (x,t).
\end{align*}
That the treatment levels are positive for all levels of the covariates gives that this quantity is well-defined. By Fubini's theorem, and using that $\mathcal T = [K]$ is discrete:
\begin{align*}
    f_{\mathbf y_i(k)}(y) &= \int_{\mathcal X}\bracks*{\sum_{t \in [K]}
    f_{\mathbf y_i(k) | t, x}(y)\prob{\mathbf t_i = t | \mathbf x_i = x}} f_{\mathbf x_i}(x)\,\text d x.
\end{align*}

By ignorability and consistency, $f_{\mathbf y_i(k) | t, x}(y) = f_{\mathbf y_i | k, x}(y)$, so:
\begin{align*}
    f_{\mathbf y_i(k)}(y) &= \int_{\mathcal X}\bracks*{\sum_{t \in [K]}
    f_{\mathbf y_i | k, x}(y)\prob{\mathbf t_i = t | \mathbf x_i = x}} f_{\mathbf x_i}(x)\,\text d x.
\end{align*}
Finally, since $f_{\mathbf y_i | k, x}(y)$ is constant with respect to $t$:
\begin{align*}
    f_{\mathbf y_i(k)}(y) &= \int_{\mathcal X}
    f_{\mathbf y_i | k, x}(y)\bracks*{\sum_{t \in [K]}\prob{\mathbf t_i = t | \mathbf x_i = x}} f_{\mathbf x_i}(x)\,\text d x \\
    &= \int_{\mathcal X}
    f_{\mathbf y_i | k, x}(y) f_{\mathbf x_i}(x)\,\text d x, \,\,\,\, \sum_{t \in [K]}\prob{\mathbf t_i = t | \mathbf x_i = x} = 1.
\end{align*}
Then:
\begin{align*}
    f_{\mathbf y_i(k)}(y) - 
    f_{\mathbf y_i(l)}(y) &= \int_{\mathcal X}
    \bracks*{f_{\mathbf y_i | k, x}(y) - f_{\mathbf y_i | l, x}(y) }f_{\mathbf x_i}(x)\,\text d x\numberthis \label{eqn:udice_helper}.
\end{align*}
By assumption 3., $f_{\mathbf y_i(k)} \neq f_{\mathbf y_i(l)}$ precisely when $f_{\mathbf y_i | k, x} \neq f_{\mathbf y_i | l, x}$ for some $x$ where $f_{\mathbf x_i}(x) > 0$.
\end{proof}
This implies the following corollary:
\begin{corollary}[Equivalence of causal conditional discrepancies and unconditional causal discrepancies]
Suppose the setup described in Setup \ref{setup}, and further make the assumptions of Lemma \ref{lem:udice}. Then a causal conditional discrepancy in equivalent to an unconditional causal discrepancy.
\end{corollary}
\begin{proof}
Follows by direct application of Lemma \ref{lem:udice} and Lemma \ref{thm:CoDiCE}, by transitivity.
\end{proof}

In this sense, we can conceptualize a 
causal discrepancy as \textit{smoothing} a causal conditional discrepancy (based on the relative contributions of a given $x \in \mathcal X$ to the weighted average, weighted by way of $f_{\mathbf x_i}(x)$). When all of the conditional discrepancies in the distributions of groups $k$ and $l$ are equal in sign (across covariate levels), a test of an unconditional causal discrepancy is equivalent to a causal conditional discrepancy.

The $k$-sample testing problem \cite{Panda2019Oct} is given by:
\begin{align}
    H_0 : F_{\mathbf y_i|k} = F_{\mathbf y_i | l} \text{ for all $k, l$}\text{   against   }H_A : F_{\mathbf y_i | k} \neq F_{\mathbf y_i | l}\text{ for some }k, l
    \label{eqn:hypo_kst}
\end{align}

The results of Lemma \ref{lem:udice} coupled with Equation \eqref{eqn:hypo_kst} suggest the following corollary, which generalizes the concept of an average treatment effect (ATE) to arbitrary $\mathcal Y$ when the effect is further \textit{identical} across covariate levels:
\begin{corollary}[Consistent $k$-sample discrepancy testing and $k$-sample causal discrepancy testing]
\label{cor:cons_discrep_causal}
Make the assumptions of Lemma \ref{lem:udice}, and further suppose that for all $x \in \mathcal X$, that:
\begin{align*}
    f_{\mathbf y_i|k}(y) - f_{\mathbf y_i|l}(y) = f_{\mathbf y_i | k, x}(y) - f_{\mathbf y_i | l, x}(y),
\end{align*}
Then a consistent test of Equation \eqref{eqn:hypo_kst} is equivalent to a test of Equation \eqref{eqn:hypo_udice}.
\end{corollary}
\begin{proof}
Starting at Equation \eqref{eqn:udice_helper} and using the assumption:
\begin{align*}f_{\mathbf y_i(k)}(y) - 
    f_{\mathbf y_i(l)}(y) &= \int_{\mathcal X}
    \bracks*{f_{\mathbf y_i | k, x}(y) - f_{\mathbf y_i | l, x}(y) }f_{\mathbf x_i}(x)\,\text d x \\
    &= f_{\mathbf y_i|k}(y) - f_{\mathbf y_i | l}(y),\,\,\,\,\int_{\mathcal X}f_{\mathbf x_i}(x)\,\text d x = 1.
\end{align*}
Therefore, $f_{\mathbf y_i(k)} \neq f_{\mathbf y_i(l)}$ precisely when $f_{\mathbf y_i | k} \neq f_{\mathbf y_i | l}$.
\end{proof}
Conceptually, this result indicates that, so long as the impact of treatment (on the outcome distribution) is identical (in magnitude) across covariate levels, then a $k$-sample discrepancy test is a consistent test for a $k$-sample causal discrepancy. \citet{Panda2019Oct} gives that a consistent test of Equation \eqref{eqn:hypo_kst} can be practically achieved via independence testing (such as through \Dcorr) assuming that further assumptions are satisfied about $F_{\mathbf y_i|k}$ and $F_{\mathbf y_i|l}$ (such as finite first and second moments).

This result is noteworthy in that it would be simple to use the intuition of Lemma \ref{thm:CoDiCE} to conclude that under causal assumptions $k$-sample discrepancy tests would be equivalent to $k$-sample causal discrepancy tests intuitively. However, the fine details of the implications of the ignorability condition reveal this to be incorrect, and we need an additional (stronger) condition (such as the one given in Corollary \ref{cor:cons_discrep_causal}) for this to be the case. Conceptually, under the condition given, the ``smoothing'' that we noted in Equation \eqref{eqn:udice_helper} need not be relevant (because the effects are all the same) to attain the proper marginal distributions for $k$-sample discrepancy testing. 

A weaker, albeit less intuitive, condition that would also give the desired result would be that $\int_{\mathcal X}f_{\mathbf y_i | k, x}(y)f_{\mathbf x_i}(x)\,\text d x = f_{\mathbf y_i | k}$. A potentially stronger condition would be unconditional ignorability $(\mathbf y_i(1), \hdots, \mathbf y_i(K)) \indep \mathbf t_i$, which coupled with positivity, can be practically achieved via randomization.

\section{Simulations}
\label{app:sims}

\subsection{Setup}

$n=100$ samples are generated when balance $\pi_b$ is high or low ($\pi_b = 0.8$ or $\pi_b = 0.4$) and dimensionality is high or low ($D = 10$ or $D=101$).

\paragraph{Covariate sampling, $2$-group}

The treatment group $\mathbf t_i \distas{iid} \Bern{\pi}$ for $i \in [n]$ with $\pi = 0.5$.

The balance of a sample $\mathbf b_i \distas{iid} \Bern{\pi_b}$ for $i \in [n]$.

\begin{figure}
    \centering
    \includegraphics[width=\linewidth]{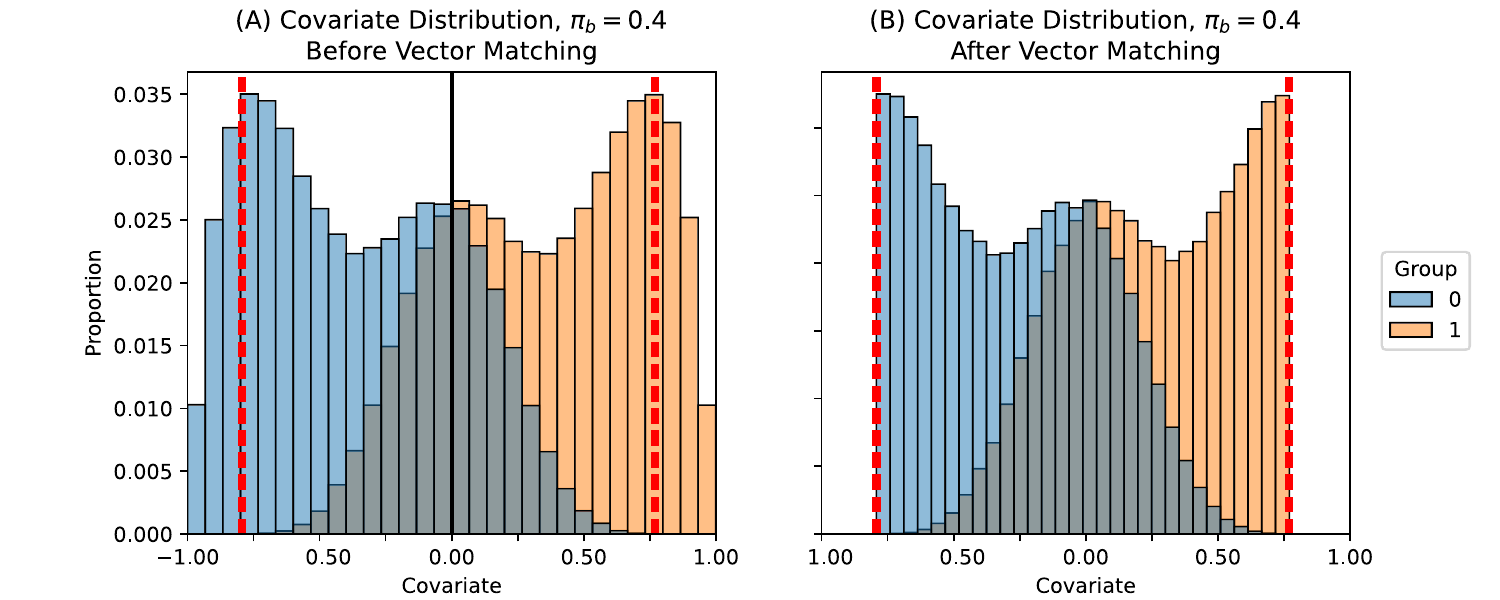}
    \caption{\textbf{Covariate Generation and Vector Matching}. \textbf{(A)} indicates the covariate distribution, with $\pi_b = 0.4$ (the \textit{low balance} regime). Simulations are conducted with $n=100$ samples, but $n=10000$ are employed here for visualization purposes. Bar heights correspond to the proportion of samples which correspond to a particular covariate range for each of the two treatment groups. The center ``hump'' is indicative of the fact that $\pi_b$ of the points have the same covariate value. Note that the two groups have distributions which are reflected about the origin $x=0$ (black line). Two vertical lines (red) indicate the ``cutoffs'' corresponding to points in group $1$ (left line) with the lowest propensity for group $1$ and points in group $1$ (right line) with the lowest propensity for group $1$. \textbf{(B)} indicates the covariate distribution after vector matching (\vm) \cite{Lopez2014}. The highest/lowest propensity scores (across groups) serve as ``cutoffs'' to filter samples so that the empirical propensity score distributions have the same support. These filtered samples are then analyzed via ``causal'' strategies for downstream inference.}
    \label{fig:vm}
\end{figure}

The values $\mathbf z_i$ are sampled independently as:
\begin{align*}
    \mathbf z_i | \mathbf b_i = b, \mathbf t_i = t \distas{d} \begin{cases}
        \Beta{10, 10} & b = 1 \\
        \Beta{2, 8} & b = 0, t = 0 \\
        \Beta{8, 2} & b = 0, t = 1
    \end{cases},
\end{align*}
and the covariate $\mathbf x_i = 2\mathbf z_i - 1$. Conceptually, $\pi_b$ of the points (ignoring the group assignments) have the same covariate distribution given by $2\Beta{10, 10} - 1$, and $1 - \pi_b$ of the points are in the right- and left-skewed distributions given by $2\Beta{2, 8}-1$ if the point is in group $0$ and $2\Beta{8, 2}-1$ if the point is in group $1$ respectively. Note further that by construction, the covariate distributions are symmetric about $0$; e.g., $f(x|0) = f(-x | 1)$ for all $x \in \mathcal X = [-1, 1]$. Figure \ref{fig:vm}(A) details the covariate generation procedure.

\paragraph{Covariate sampling, $K$-group}

The treatment probability vector $\vec \pi$ is:

\begin{align*}
    \pi_k = \begin{cases}
        \pi & k = 1 \\
        \frac{1 - \pi}{K - 1} & k > 1
    \end{cases}.
\end{align*}
The treatment group $\mathbf t_i \distas{iid} \Categorical{\vec \pi}$, where $\Categorical{\vec \pi}$ is the categorical distribution (e.g., $\prob{\mathbf t_i = k} = \pi_k$ for $k \in [K]$ when $\vec \pi$ is in the $K$-probability simplex). Conceptually, $\pi$ is the fraction of points in the first group, and $1 - \pi$ is the fraction of points that are evenly distributed amongst the remaining $K-1$ groups.

The balance of a sample $\mathbf b_i \distas{iid} \Bern{\pi_b}$ for $i \in [n]$.

The values $\mathbf z_i$ are sampled independently as:
\begin{align*}
    \mathbf z_i | \mathbf b_i = b, \mathbf t_i = t \distas{d} \begin{cases}
        \Beta{10, 10} & b = 1 \\
        \Beta{2, 8} & b = 0, t = 1 \\
        \Beta{8, 2} & b = 0, t \neq 1
    \end{cases},
\end{align*}
and the covariate $\mathbf x_i = 2\mathbf z_i - 1$. Conceptually, $\pi_b$ of the points have the same distribution given by $2\Beta{10, 10} - 1$, and $1 - \pi_b$ of the points assigned to group $1$ are in the right-skewed distribution given by $2\Beta{2, 8} - 1$, with $1 - \pi_b$ of the points in the groups $\{2, ..., K\}$ in the left-skewed distribution given by $2\Beta{8,2}-1$ (the first group is right-skewed, and the remaining groups have the same left-skewed distribution). Note that by construction, for all $t' \in \{2, ..., K\}$, that $f(x | 1) = f(-x | t')$, by a similar argument to above.

\paragraph{Common characteristics and the signal to noise ratio} Several of the below simulations use the non-linear (but monotonic) sigmoid function, which is defined for $x \in \mathbb R$ as:
\begin{align*}
    \sigmoid(x) &= \frac{1}{1 + \exp(-x)} = \frac{\exp(x)}{\exp(x) + 1} \in [0, 1]
\end{align*}

Further, all of the simulations will control the signal to noise ratio in successive dimensions. Intuitively, if the level of signal is not decreased (per dimension) as the dimensionality increases, the signal to noise ratio of the simulation will increase to infinity. The vector $\vec \beta \in \mathbb R^{D}$ for a given dimensionality $D \in \natn$ is given by:
\begin{align*}
    \beta_p = \frac{2}{p^{q}}
\end{align*}
where $q > 1$. Conceptually, $\beta_p$ controls the amount of signal for the outcome in a given dimension $p \in [D]$, and $\beta_{p + 1} < \beta_p$ for all $p \in [D]$ (higher dimensions have \textit{less} signal). 

The noise in a given simulation is $\vec{\pmb \epsilon}_{i} \distas{iid} \Norm[D]{\vec 0, \frac{1}{4}I_D}$ unless otherwise indicated. Since the noise $\vec{\mathbf \epsilon}_{i}$ has variance $1$ for all dimensions $p \in [D]$, note that the SNR for all contexts except for \textit{Heteroskedastic} is, where the signal $\mathbf s_i$:
\begin{align*}
    SNR &= \frac{\expect{||\mathbf s_i||^2_2}}{\expect{||\pmb \epsilon_i||_2^2}} \\
    &\leq \frac{\norm{\vec\beta}_2^2}{\norm{\vec{\pmb \epsilon}_{i}}_2^2},\,\,\,\,\Delta \leq 1 \\
    &= \frac{2}{D}\sum_{p = 1}^D \frac{1}{p^x}.
\end{align*}
The right-most quantity is a $p$-series, and therefore converges for $x > 1$. Therefore, the SNR is finite for large $D$, and converges to $0$ as $D \rightarrow \infty$. That the SNR is finite for the heteroskedastic simulation is discussed in its respective paragraph.

\paragraph{Simulations are rotated to ensure tests are incorporating information across dimensions} To ensure the flexibility of the described techniques to high dimensional investigations so that no simulations are benefiting from only looking at the first dimension (which contains a much higher quantity of signal than successive dimensions), the outcomes are rotated. For all simulations, the outcome is $\mathbf R\vec{ \mathbf y_i}$, where $\mathbf R \distas{d} \text{Haar}_D$ is a Haar random orthogonal matrix. Conceptually, realizations $R$ of $\mathbf R$ are $D \times D$ rotation matrices drawn from the Haar distribution, which is the uniform distribution on special orthogonal matrices of dimension $D$ (SO$(D)$) \cite{scipy,haar}. 

\subsection{Simulation Settings}

\begin{figure}[h!]
    \centering
    \includegraphics[width=\linewidth]{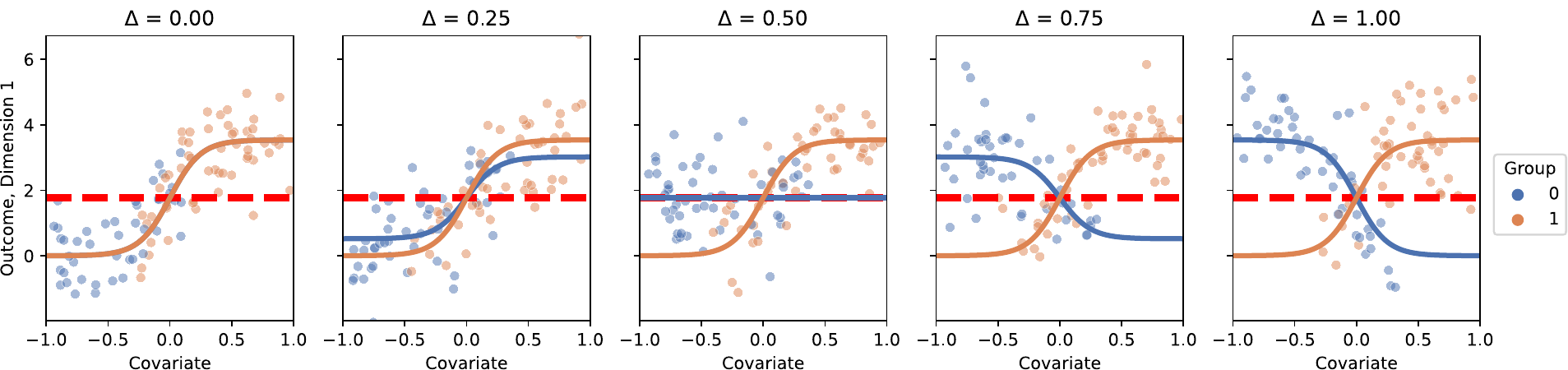}
    \caption{\textbf{Sigmoidal causal conditional discrepancy simulation across range of effect sizes}. The group-specific outcomes at each covariate level, for dimension $1$ ($D = 10$). The horizontal line (red) corresponds to the line $\frac{5}{2\sqrt 2}$. The solid line (orange) indicates the average outcome for group $1$ at a particular covariate level. The average outcome for group $0$ at a particular covariate level are rotated about the horizontal line (red) by $\Delta \pi$ radians.  At $\Delta = 1$, the outcome distributions are identical unconditional on the covariates because the covariate distribution of Group $0$ is a reflection of that of Group $1$. $n=100$ samples are then generated for each of the two groups (dots) according to the $2$-group covariate sampling procedure described by simulating a covariate, identifying the average outcome for the assigned group of the sample at the sample's covariate level, and then adding sample-specific noise (independent and identically distributed across all samples).}
    \label{fig:sigmoidal}
\end{figure}

\paragraph{Sigmoidal} Let $q = 1.5$. Conceptually, consider a plot of the outcome (per dimension) relative the covariate value, as in Figure \ref{fig:sigmoidal}. the outcome distribution (for a given dimension) is \textit{rotated} by $\Delta\pi$ radians (the \textit{effect size}, where $\Delta$ ranges from a minimum of $0$ to a maximum of $1$) for one group relative the other about the horizontal line with an intercept at $\frac{5}{p^{1.5}}$. This corresponds to the value attained by the sigmoid at the \textit{midpoint} $x = 0$ in the covariate distributions. That the covariate distributions are effectively \textit{reflections} of one another about the point $x = 0$ gives that there is no unconditional causal discrepancy when the effect size $\Delta = 1$.

The rotation factor $r = \cos(\Delta\pi)$ rotates the outcomes by $\Delta\pi$ radians about the horizontal line at $y = 0$. The outcomes are:
\begin{align*}
    \vec{\mathbf y}_i(t)| x &= \begin{cases}
        \left(r\left(5\sigmoid(8x) -  \frac{5}{2}\right) + \frac{5}{2}\right)\vec \beta + \vec{\pmb\epsilon}_i ,& t = 0 \\
        5\sigmoid(8x)\vec \beta + \vec {\pmb\epsilon}_i ,& t = 1
    \end{cases}
\end{align*}
By construction, note that $\vec{\mathbf y}_i(0)| x \eqdist \vec{\mathbf y}_i(1)| -x$ when $\Delta = 1$, and further, that $f(x | 1) = f(-x | 0)$. Therefore, $F_{\mathbf y_i(1)} = F_{\mathbf y_i(0)}$ unconditionally when $\Delta = 1$, and no unconditional causal discrepancy from Definition \ref{def:udice} exists. Figure \ref{fig:sigmoidal} illustrates the sigmoidal simulations from $\Delta = 0.0$ to $\Delta = 1.0$.

\begin{figure}[h!]
    \centering
    \includegraphics[width=\linewidth]{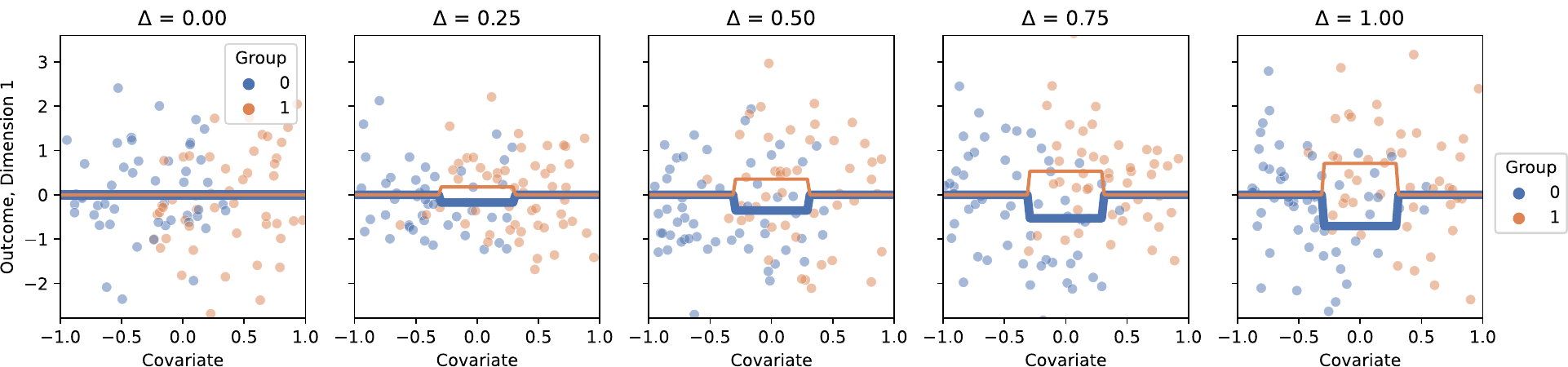}
    \caption{\textbf{Non-monotone causal conditional discrepancy simulation across range of effect sizes}. The group-specific outcomes at each covariate level, for dimension $1$ ($D = 10$). Group-specific average outcomes are shown (solid lines). The causal conditional discrepancy can be characterized by the $2\Delta \vec\beta$ difference between the two groups for covariates in $[-0.3, 0.3]$. $n=100$ samples are then generated for each of the two groups (dots) according to the $2$-group covariate sampling procedure described by simulating a covariate, identifying the average outcome for the assigned group of the sample at the sample's covariate level, and then adding sample-specific noise (independent and identically distributed across all samples).}
    \label{fig:nm}
\end{figure}

\paragraph{Non-monotone} Let $q = 1.5$. Conceptually, There is only a covariate-specific effect for points when $\mathbf x_i \in [-.3, .3]$. This effect is $\Delta\vec\beta$ for points that are in group $1$, and $-\Delta \vec\beta$ for points in group $2$ ($\Delta$ is the \textit{effect size} ranging from $0$ to $1$). The outcomes are:
\begin{align*}
    \vec{\mathbf y}_i(t)| x &= \begin{cases}
        \vec{\pmb \epsilon}_i ,& x \not\in [-0.3, 0.3] \\
        \Delta \vec \beta + \vec{\pmb \epsilon}_i ,& t = 1, x \in [-0.3, 0.3] \\
        -\Delta \vec \beta + \vec{\pmb \epsilon}_i ,& t = 0, x \in [-0.3, 0.3]
    \end{cases}.
\end{align*}
For a given $x \in \mathcal X = [-1, 1]$, the covariate-specific effect is:
\begin{align*}
    \mathbf y_i(1) - \mathbf y_i(0)|x &= \begin{cases}
        0 ,& x \not\in [-0.3, 0.3] \\
        2\Delta \vec\beta ,& x \in [-0.3, 0.3]
    \end{cases},
\end{align*}
So a causal conditional discrepancy exists when $\Delta > 0$. Note that by using ignorability and positivity (which are true, by construction) and that the distributions are equal for $x \not\in [-0.3, 0.3]$:
\begin{align*}
    \expect{\mathbf y_i(1) - \mathbf y_i(0)} &= \Delta \vec\beta \bracks*{\int_{-0.3}^{0.3}f(x|0)\text d x + \int_{-0.3}^{0.3}f(x | 1)\text dx} \\
    &= 2\Delta \vec \beta \int_{-0.3}^{0.3}f(x | 0)\text dx \\
    &\begin{cases}
        > 0, & \Delta > 0 \\
        = 0, & \Delta = 0
    \end{cases},
\end{align*}
where the second to last line follows because $f(x|0) = f(-x|1)$ by construction, and the intervals are symmetric of the form $[-u, u]$ (for instance, using the substitution $u=-x$ for the right-most quantity gives the desired result). The last line follows because the interval $[-0.3, 0.3]$ is in the support of $\mathbf x_i$ conditional on the group $t$. Since the expectations are unequal, the distributions $F_{\mathbf y_i(1)} \neq F_{\mathbf y_i(0)}$. This shows that an unconditional causal discrepancy from Definition \ref{def:udice} also exists when $\Delta > 0$.  Figure \ref{fig:nm} illustrates the non-monotonic simulations from $\Delta = 0.0$ to $\Delta = 1.0$.

\begin{figure}[h!]
    \centering
    \includegraphics[width=\linewidth]{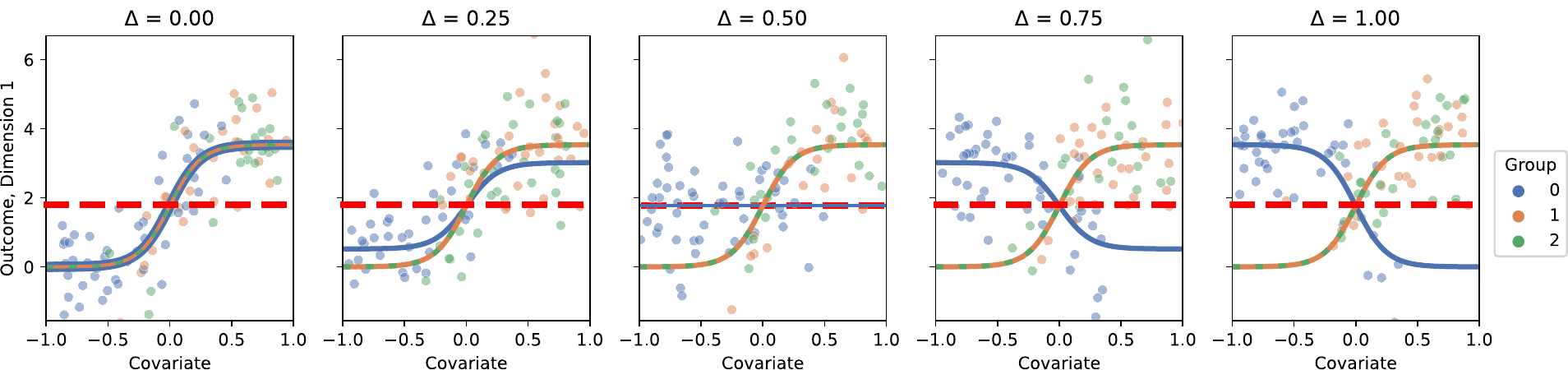}
    \caption{\textbf{$K$-Group causal conditional discrepancy simulation across range of effect sizes}. The group-specific outcomes at each covariate level, for dimension $1$ ($D = 10$). As in Figure \ref{fig:sigmoidal}, group $0$ is rotated about the line $\frac{5}{2}D^{-\frac{1}{4}}$ for increasing $\Delta$. At $\Delta = 1$, the outcome distributions are identical unconditional on the covariates because the covariate distribution of Group $0$ is a reflection of that of Group $1$ or $2$. $n=100$ samples are then generated for each of the two groups (dots) according to the $2$-group covariate sampling procedure described by simulating a covariate, identifying the average outcome for the assigned group of the sample at the sample's covariate level, and then adding sample-specific noise (independent and identically distributed across all samples).}
    \label{fig:kclass}
\end{figure}

\paragraph{$K$-group} Let $q = 1.1$. This simulation is conceptually identical to the Sigmoidal simulation. With $r = \cos(\Delta \pi)$ the rotation factor, the outcomes are:
\begin{align*}
    \vec{\mathbf y}_i | \mathbf t_i = t, \mathbf x_i = x &= \begin{cases}
        \left(r\left(5\sigmoid(8\mathbf x_i) -  \frac{5}{2}\right) + \frac{5}{2}\right)\vec \beta + \vec \epsilon_i ,& \mathbf t_i = 1 \\
        5\sigmoid(8\mathbf x_i)\vec \beta + \vec \epsilon_i ,& \mathbf t_i > 1
    \end{cases}.
\end{align*}
Again, when the effect size is at a maximum of $\Delta = 1$, the covariate distribution for group $1$ is a \textit{reflection} of the covariate distribution for groups $\{2, ..., K\}$ about the point $x=0$. Therefore there is no unconditional causal discrepancy. Figure \ref{fig:kclass} illustrates the $K$-group simulations from $\Delta = 0.0$ to $\Delta = 1.0$.

\begin{figure}[h!]
    \centering
    \includegraphics[width=\linewidth]{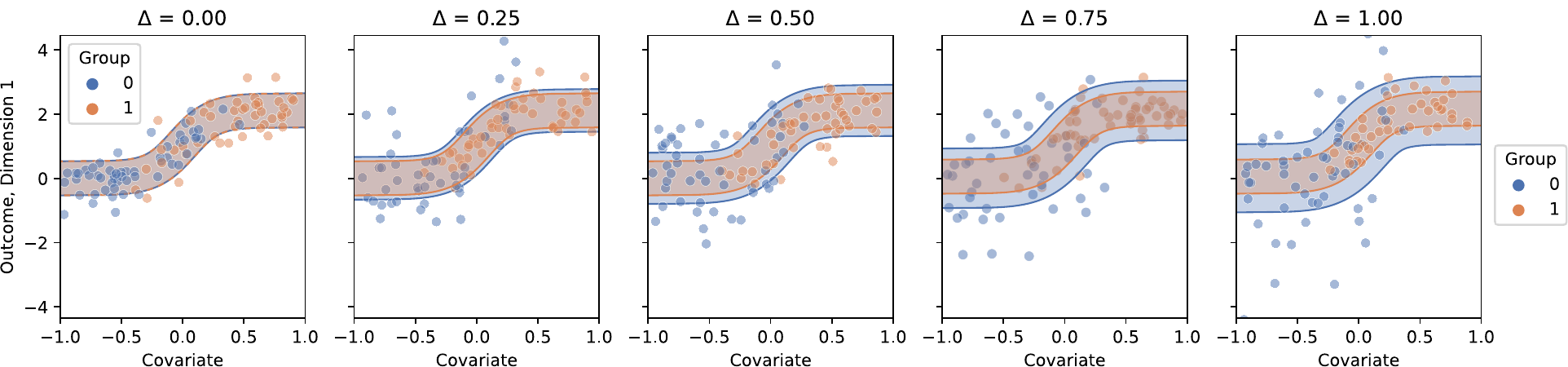}
    \caption{\textbf{Heteroskedastic causal conditional discrepancy simulation across range of effect sizes}. The group-specific outcomes at each covariate level, for dimension $1$ ($D = 10$). The group-specific mean outcome at a particular covariate level is identical between groups $0$ and $1$. The causal effect is captured by a heteroskedastic covariance model (shaded boxes indicate $\pm 1$ standard deviation from the mean). The effect size $\Delta$ indicates the order of magnitude that the standard deviation from the mean is larger for group $0$ than that of group $1$. At $\Delta = 1.0$, the standard deviation from the mean at a particular covariate level for group $0$ is double that of group $1$.}
    \label{fig:hetero}
\end{figure}

\paragraph{Heteroskedastic} Let $q = 1.5$. Conceptually, an unconditional causal discrepancy is present, in that the difference between $F_{y_i(1) | x} = F_{y_i(2)|x}$ can be characterized by $\mathbf y_i(1)$ having a larger covariance than $\mathbf y_i(2)$ (and the difference does not depend on $x$). The outcome model is:
\begin{align*}
    \vec{\mathbf y}_i \cond \mathbf t_i = t, \mathbf x_i = x &= \begin{cases}
        5\sigmoid\parens*{8\mathbf x_i}\vec \beta + \sqrt{1 + \Delta}\pmb{\epsilon}_i & \mathbf t_i = 0 \\
        5\sigmoid\parens*{8\mathbf x_i}\vec \beta + \pmb{\epsilon}_i & \mathbf t_i = 1
    \end{cases}
\end{align*}
for all $x \in \mathcal X$, indicating that an \textit{unconditional} causal discrepancy exists when $\Delta > 0$. Figure \ref{fig:hetero} illustrates the heteroskedastic simulations from $\Delta = 0.0$ to $\Delta = 1.0$. By construction, for any $D$, the SNR is $\Delta\sum_{p = 1}^D \frac{2}{p^q}$, so the SNR is finite for any $D$ for $\Delta \leq 1$.

\section{Multinomial Regression Model} For our proposed inferential procedure, we use the baseline-category logit model for vector matching (\vm) \cite{Lopez2014}. With $\vec x_i$ the vector of covariates and group $k$ is arbitrarily the baseline, the model is:
\begin{align*}
    \log\parens*{\frac{r(l, x_i)}{r(k, x_i)}} &= \vec \beta_l^\top \vec x_i,
\end{align*}
where $r(l, x_i)$ is the generalized propensity score of item $i$ in group $l$. The generalized propensity scores are:
\begin{align*}
    r(l, x_i) &= \begin{cases}
    \frac{\exp\parens*{\vec \beta_l^\top \vec x_i}}{1 + \sum_{t \neq k}\exp\parens*{\vec \beta_t^\top\vec x_i}},& l \neq k \\
    \frac{1}{1 + \sum_{t \neq k}\exp\parens*{\vec \beta_t^\top\vec x_i}},& l = k 
    \end{cases} \numberthis \label{eqn:multinom}
\end{align*}
 The model is fit using the \sct{statsmodels} package \cite{seabold2010statsmodels} in the \sct{python} programming language to obtain estimates of the regression coefficients $\hat {\vec\beta}_l$ for all $l \neq k$. Estimated propensity scores $\hat r(l, x_i)$ are obtained for all samples $i$ and for all groups $l \in [K]$ by plugging in the estimated regression coefficients $\hat {\vec\beta}_l$ for $l \neq k$ to the Equations given in \eqref{eqn:multinom}. In the univariate regime, \vm~corresponds to identifying the highest/lowest propensity samples within a given treatment group for all treatment groups, finding the smallest/largest across all treatment groups for a given treatment group, and finally filtering points using the identified cutoffs, as illustrated in Figure \ref{fig:vm}(B) and according to Equation \eqref{eqn:vm}.

\end{document}